\documentclass[conference]{IEEEtran}
\IEEEoverridecommandlockouts

\usepackage{mathptmx} 
\usepackage{cite}
\usepackage{amsmath,amssymb,amsfonts}
\usepackage{algorithmic}
\usepackage{graphicx}
\usepackage{textcomp}
\usepackage{xcolor}
\usepackage{fancyhdr}
\usepackage[hyphens]{url}

\def\BibTeX{{\rm B\kern-.05em{\sc i\kern-.025em b}\kern-.08em
    T\kern-.1667em\lower.7ex\hbox{E}\kern-.125emX}}

\usepackage[keeplastbox]{flushend}
\usepackage[bookmarks=true,breaklinks=true,letterpaper=true,colorlinks,linkcolor=black,citecolor=blue,urlcolor=black]{hyperref}

\usepackage{graphicx}
\usepackage{epstopdf}
\usepackage{setspace} 
\usepackage{subfig}
\usepackage[ruled,vlined,linesnumbered]{algorithm2e}
\usepackage{url}
\usepackage{amsmath}
\usepackage{pifont}
\usepackage{fixltx2e}
\usepackage{anyfontsize}
\usepackage{kantlipsum}
\usepackage{multirow}
\usepackage[font=bf, skip=1pt]{caption}
\usepackage{array}
\usepackage{comment}
\usepackage{mwe}
\usepackage{enumitem}

\usepackage{tikz}
\usepackage{kantlipsum}

\newcommand*\whitecircled[1]{\tikz[baseline=(char.base)]{
            \node[shape=circle,fill=black,font=\bfseries,inner sep=1pt] (char) {\textcolor{white}{#1}};}}
\newcommand*\redcircled[1]{\tikz[baseline=(char.base)]{
            \node[shape=circle,fill=red,font=\bfseries,inner sep=1pt] (char) {\textcolor{white}{#1}};}}

\iftrue
\newcommand{\nskim}[1]{{\color{red}[\textbf{\sc nskim}: \textit{#1}]}}
\newcommand{\mj}[1]{{\color{blue}[\textbf{\sc mj}: \textit{#1}]}}
\newcommand{\jie}[1]{{\color{orange}[\textbf{\sc jie}: \textit{#1}]}}
\else
\newcommand{\nskim}[1]{}
\newcommand{\mj}[1]{}
\newcommand{\jie}[1]{}
\fi

\newcommand{\mycomment}[1]{}
\newcommand{\ignore}[1]{}

\usepackage{color}
\usepackage{soul}
\usepackage{xspace}

\newcommand{\newedit}[1]{#1\color{black}\xspace}

\begin{document}

\title{Revamping Storage Class Memory With Hardware Automated Memory-Over-Storage Solution}
\author{
{\large Jie Zhang$^{1}$, Miryeong Kwon$^{1}$, Donghyun Gouk$^{1}$, Sungjoon Koh$^{1}$, Nam Sung Kim$^{2}$} \\
\vspace{-2pt}
{\large Mahmut Taylan Kandemir$^{3}$, Myoungsoo Jung$^{1}$} \\
\vspace{-2pt}
       {\large \emph{Computer Architecture and Memory Systems Laboratory,}}\\
\vspace{-2pt}
       {\normalsize{Korea Advanced Institute of Science and Technology (KAIST)$^{1}$, University of Illinois Urbana-Champaign$^{2}$}}\\
\vspace{-3pt}
       {\normalsize{Pennsylvania State University$^{3}$}}\\
\vspace{-2pt}
	   {\large {http://camelab.org}}
       }



\maketitle

\begin{abstract}
Large persistent memories such as NVDIMM have been perceived as a disruptive memory technology, because they can maintain the state of a system even after a power failure and allow the system to recover quickly.
However, overheads incurred by a heavy software-stack intervention seriously negate the benefits of such memories. 
First, to significantly reduce the software stack overheads, we propose HAMS, a hardware automated Memory-over-Storage (MoS) solution.
Specifically, HAMS aggregates the capacity of NVDIMM and ultra-low latency flash archives (ULL-Flash) into a single large memory space, which can be used as a working memory expansion or persistent memory expansion, in an OS-transparent manner. 
HAMS resides in the memory controller hub and manages its MoS address pool
over conventional DDR and NVMe 
interfaces; it employs a simple hardware cache to serve all the memory requests from the host MMU after mapping the storage space of ULL-Flash to the memory space of NVDIMM. 
Second, to make HAMS more energy-efficient and reliable, we propose an ``advanced HAMS'' which removes unnecessary data transfers between NVDIMM and ULL-Flash after optimizing the datapath and hardware modules of HAMS. 
This approach unleashes the ULL-Flash and its NVMe controller from the storage box and directly connects the HAMS datapath to NVDIMM over the conventional DDR4 interface. 
Our evaluations show that HAMS and advanced HAMS can offer 97\% and 119\% higher 
system performance than a software-based NVDIMM design, 
while costing 41\% and 45\% lower energy, respectively.

\end{abstract}

\section{Introduction}
\label{sec:intro}
Recently, persistent memories such as PRAM \cite{lee2009} and 3D XPoint \cite{xpoint2015} have received 
a considerable attention as their non-volatile intrinsic, high density and low power consumption can benefit modern datacenters and high-performance computers. 
For such systems, back-end storage is required for recovering from system failures and crashes. 
Since the persistent memories can spontaneously and instantaneously recover all memory states, 
they can eliminate a large number of accesses to the back-end storage and associated runtime overheads \cite{narayanan2012whole, volos2011mnemosyne, kolli2016high}. Besides, enterprise workstations and servers employ the persistent memory with DirectAccess (DAX) \cite{zdnet-dax,hpe-dax}, which brings the advantages of unprecedented levels of performance and data resiliency \cite{snia-dax}. 

There are three standard persistent memory types (i.e., NVDIMM-N/F/P). NVDIMM-F directly integrates flash into a dual-inline memory module (DIMM) to provide a high capacity similar to storage. However, NVDIMM-F cannot simply replace DRAM, as it only exposes a block interface. \newedit{NVDIMM-P such as Optane DC PMM is byte-addressable, but its performance using the app direct mode to support data persistence is yet 6$\times$ worse than DRAM \cite{optane2017intel,optane-perf}.} In contrast, NVDIMM-N aims to offer ``byte-addressable'' persistency with DRAM-like performance \cite{sainio2016nvdimm}. 
NVDIMM-N generally consists of a small flash device and multiple DRAM modules with a battery. NVDIMM-N can be useful for a wide range of data-intensive applications such as database management system (DBMS) \cite{arulraj2017build}, transaction processing \cite{wang2014scalable,liu2017dudetm}, data analytics \cite{chen2016bridging}, and checkpointing \cite{gao2015real}. However, the memory space of NVDIMM-Ns 
(\textit{e.g.}, 4GB $\sim$ 64GB) is 
considerably smaller than that of NVDIMM-P and persistent storage 
devices such as solid state drives (SSDs). Furthermore, the capacity of DRAM in NVDIMM-Ns  is 
constrained by poor scaling of battery that needs to supply the power for DRAM backup operations when a power failure occurs 
\cite{zhao2012optimizing,bhati2013coordinated}. For example, for the past two decades, the storage density of DRAM has increased by many orders of magnitude, whereas the energy density of lithium-ion battery has only tripled \cite{kateja2017viyojit}.


\begin{figure}
	\centering
	\includegraphics[width=1\linewidth]{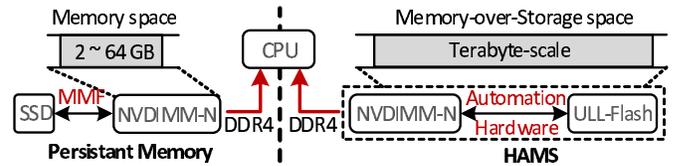}
	\vspace{-20pt}
	\caption{\label{fig:summary}NVDIMM-N vs. HAMS.\vspace{-15pt}}
	\vspace{-10pt}
\end{figure}


A possible solution to build a large and scalable, yet persistent memory space is to use NVDIMM-N together with SSD and memory-mapped files (MMFs), 
which can be implemented in an OS memory manager or a file system. 
This allows data-intensive applications to access a large storage space using conventional load/store instructions. 
However, we observe that such MMF-assisted persistent memory can degrade the application performance at the user-level by 48\%, on average, compared to an NVDIMM-N-only solution (cf. Section \ref{sec:sw-mem-expansion}). 
Such severe performance degradation is caused by not only the long stall latency in accessing SSD but also the software overhead and
frequent data copies between the user and system memory spaces in a conventional storage stack.

\newedit{Tackling the aforementioned limitations, we propose HAMS, a \underline{\textbf{H}}ardware \underline{\textbf{A}}utomated \underline{\textbf{M}}emory-over-\underline{\textbf{S}}torage (MoS) solution that aggregates 
the memory capacity of NVDIMM-N and the storage capacity of new ultra-low latency flash archives, referred to as \emph{ULL-Flash} \cite{koh2018exploring, zhang2018flashshare}, into a single large memory space (\textit{cf.} Figure \ref{fig:summary}). }
The large monolithic memory space of HAMS can be used as a working memory or a persistent memory expansion. 
Our HAMS resides in the memory controller hub, and manages its MoS address pool by leveraging the conventional DDR4 and NVMe 
interfaces. 
To this end, HAMS employs a simple hardware cache to handle all the memory requests from the host memory management unit (MMU) by mapping the storage space of ULL-Flash to the memory space of NVDIMM-N. 
In case of an NVDIMM-N cache miss, HAMS internally manages the NVMe commands and I/O request queues while hiding all the NVMe protocol and interface management overheads from the OS, such that data requested by MMU are always served by NVDIMM-N.

While the ``baseline'' design of HAMS can offer a 
20GB/s peak bandwidth, it can still 
yield sub-optimal system performance, especially when running large-scale data-intensive applications 
due to some inefficiencies, described subsequently.
First, handling NVDIMM-N cache misses requires data transfers between NVDIMM-N and ULL-Flash. That is, HAMS needs to go through both DDR4 and PCIe interfaces, including physical layers, controllers and protocol managers, to handle NVDIMM-N cache misses. 
However, the PCIe bandwidth is insufficient to expose the full potential of ULL-Flash to HAMS.
Consequently, the data transfers to handle frequent NVDIMM-N cache misses for large data-intensive applications can contribute to as high as 47\% of the total memory access latency of HAMS.
Second, some data may redundantly exist in the internal DRAM of both NVDIMM-N and ULL-Flash, used as cache and/or buffer. 
For example, most modern SSDs, including ULL-Flash, employ 
large internal DRAMs,
to buffer/cache all incoming I/O requests to hide the long latency of the underlying flash.  
This would help SSDs improve performance when employed in a block-storage file system, 
but it wastes power and increases the internal complexity of SSDs 
when employed 
for a MoS-based solution.

To address these limitations, 
we also propose to aggressively integrate
HAMS into existing computer systems by 
modifying its datapath and hardware modules.
This makes the baseline solution more energy-efficient and reliable, as far as data persistency is concerned. 
This ``advanced HAMS'' unleashes ULL-Flash and its NVMe controller from the storage box and directly connects their datapath to NVDIMM-N. 
To this end, we propose to slightly modify the NVMe controller within ULL-Flash, by 
incorporating a new register-based interface 
and tightly integrating the interface with the DDR4 interface of HAMS.
This aggressive integration allows ULL-Flash to access the DRAM 
devices in NVDIMM-N without any intervention from HAMS, and removes the DRAM buffer from ULL-Flash while enabling full NVMe functionality.

Our evaluation results show that HAMS and advanced HAMS provide 97\% and 119\% higher system performance, than the MMF-based NVDIMM-N+SSD hybrid design, while consuming 41\% and 45\% less system energy, respectively.




\ignore{
While future DRAM capacity is limited at XX nm, inexpensive, reliable and large storage capacities are still desirable in almost all application domains. As an alternative option, ultra-low latency (ULL) storage that leverages new memory technologies such as 3D X-Point and storage class memory (SCM) can be integrated into memory hierarchy and take over a certain part of current DRAM roles. For example, Intel Optane memory offers XX us and XX us for reads and writes, repsectively, which can be feasible to accelerate response time of diverse applications in the range of XX and XX time \cite{optane2017intel}. Similarly, Flash is in innovation as SCM to provide shorter latency and better reliability. ULL-Flash that uses new flash memory shortens the conventional flash-based storage by XX times, thereby providing XX us and XX us for reads and writes, perspectively.

These ULL storage devices open a new door to change memory hierarchy, but still several challenges exist to be used as working memory. First, the I/O granularity of their device-level operations is a logical block or page just like conventional solid state drives (SSDs). Second, the physical interface and protocol of ULL storage still leverage PCI Express (PCIe) and non-volatile memory express (NVMe), respectively, and for the host applications to access over byte access granularity, they need to assistance of kernel 3rd party software I/O library's assistance for the byte accesses, and the applications will face much longer latency than the current DRAM grantees.
}

\ignore{
In this paper, we propose HAMS, a large-scale working memory that bends DRAM (as a form of NVDIMM-N) with ULL storage, which can overcome the scaling limit as well as performance degradation that brought by DRAM and SSD respectively.
In contrast to a conventional heterogeneous system that uses memory-mapped file I/O or kernel library to support byte-addressability on storage with DRAM, HAMS automates memory caching and SSD interface management, which expose a large space of working memory pool.

The one of challenges behind HAMS design is that, while there are multiple options that we need to explore from software and hardware perspectives, there is no simulation framework that can capture hardware details from CPU to level-level NVM media as well as be feasible to capture software-side effects with our fine hardware tuning. Thus, we firstly model ULL storage that can explore different types of NVM and SSD internal hardware architecture with high fidelity, and integrate it to a full system simulator, gem5 \cite{binkert2011gem5}, by revising the datapath from memory controller and I/O bridge\footnote{All source codes of the full system simulation that integrates high-fidelity SSD storage model are available to freely download in a public domain}. This proposed simulation framework enables us to execute data-intensive applications on a real Linux system, while examining full design spaces on the datapath from the top to the bottom.

We then implement two possible HAMS options on the framework: i) loosely-coupled HAMS (LHAMS) and ii) tightly-coupled HAMS (THAMS). While both LHAMS and THAMS cache the memory requests in DRAM and provide an unified memory address space to CPU by hiding the complexity of SSD accesses,  they manage DRAM and ULL SSD over different datapath and hardware designs. LHAMS leverages existing interfaces and protocols for both DRAM and NVMe ULL SSD, which no needs to modify the underlying memory and storage modules. In contrast, THAMS is a more aggressive design that allows SSD controller can directly access DRAM controller with a minor hardware revision, which can remove external data movements on the north-bridge side.
}

\section{Background}
\label{sec:background}
In this section, we first describe the key hardware components of persistent memory and the storage stack for heterogeneous memory expansion. 
Next, we give the hardware and firmware details of ULL-Flash. 

\begin{figure}
	\centering
	\includegraphics[width=1\linewidth]{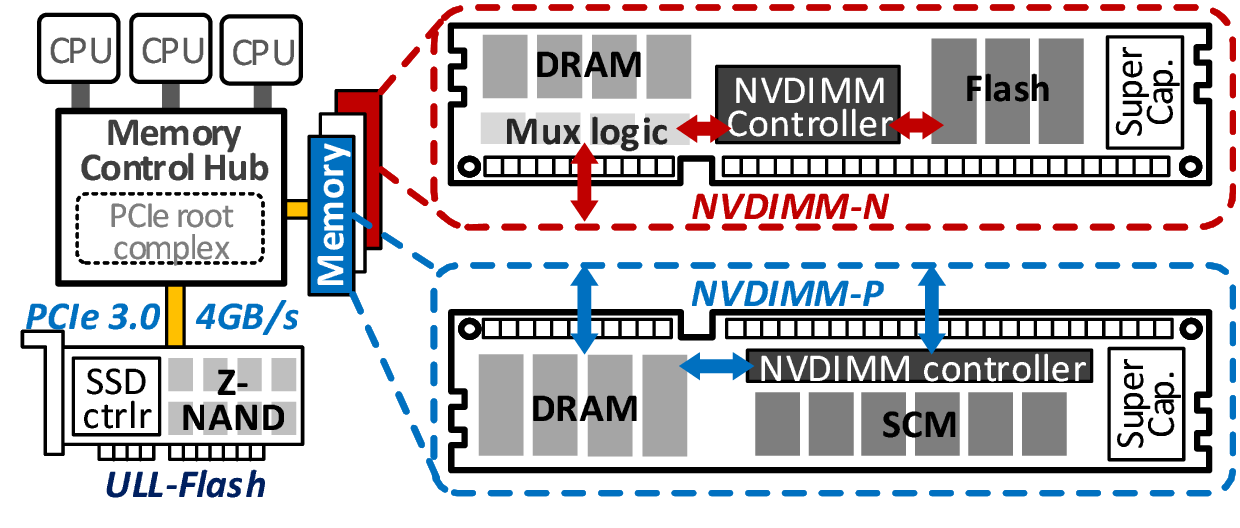}
	\vspace{-15pt}
	\caption{\label{fig:bck_persist}Persistent memory and storage.\vspace{-10pt}}
\end{figure}

\subsection{Persistent Memory and Storage}
\label{sec:persist-mem}
Figure \ref{fig:bck_persist} depicts a high-level view of a system architecture that includes NVDIMM-N/P and ULL-Flash. 
NVDIMM-N/P is attached to the memory controller hub (MCH) through a  
DDR memory bus, and operates as a standard Registered DIMM (RDIMM) for the CPU, whereas ULL-Flash is connected to MCH via a PCIe root complex as block storage.  

\begin{table}[]
\resizebox{\linewidth}{!}{
\begin{tabular}{|l|c|c|c|c|}
\hline
\textbf{Types} & \textbf{Capacity} & \textbf{OS intervention} & \textbf{Performance} & \textbf{Byte-addressable} \\ \hline
\textbf{NVDIMM-N}\cite{jesd248} & Low & No & DRAM-like & Yes \\ \hline
\textbf{NVDIMM-F}\cite{sainio2016nvdimm} & High & Yes & Slow & No \\ \hline
\textbf{NVDIMM-P}\cite{guddekoppa2016method} & Medium & Yes & Medium & Yes \\ \hline
\textbf{HAMS} & {\color{red}\textbf{High}} & {\color{red}\textbf{No}} & {\color{red}\textbf{DRAM-like}} & {\color{red}\textbf{Yes}} \\ \hline
\end{tabular}}
\caption{Feature comparison across different persistent memories and HAMS. \label{tab:feature}}
\vspace{-20pt}
\end{table}

\noindent \textbf{Persistent memory.}
There are three standard incarnations of persistent memory, 
NVDIMM-N \cite{jesd248}, -F \cite{sainio2016nvdimm} and -P \cite{guddekoppa2016method}. 
Table \ref{tab:feature} summarizes the key differences between these three types of persistent memory and our design.
NVDIMM-N is a JEDEC standard for a persistent memory module, which includes DRAM devices, a supercapacitor, multiplexers, and a small flash device. The supercapacitor is used as an energy source for the DRAM backup operations when a power failure occurs. 
The multiplexers are located between the DRAM and the standard DIMM connector to a memory bus, and they isolate the DRAM from the memory bus when backup and restore operations take place.
The flash, as a backup storage medium, has the same capacity as the DRAM, and it is invisible to users. While the host directly accesses the DRAM of NVDIMM-N, its controller internally migrates DRAM data to flash upon a power failure and this migration typically takes tens of seconds \cite{jesd248}. The controller restores the data from flash to the DRAM on the next boot, thereby providing non-volatility. 
In contrast, NVDIMM-F consists of multiple flashes without DRAM. Since it is normally used as block storage, NVDIMM-F requires both file system and OS support, similar to conventional SSDs. NVDIMM-P combines the design strategies of NVDIMM-N and NVDIMM-F, and employs a byte-addressable interface. 
However, NVDIMM-P such as Optane DC PMM exhibits 6$\times$ lower performance than DRAM, and does not allow direct access to its internal DRAM as well as requires OS-level support to enable persistent memory accesses. This by far makes NVDIMM-N the only persistent memory that supports DRAM performance with the byte-addressability. 
Considering this aspect, in this paper, we use the terms ``NVDIMM" and ``NVDIMM-N" interchangeably.


\noindent \textbf{Storage.}
All the high-performance SSDs, including ULL-Flash, are connected to another part of the MCH, PCIe root complex. 
A PCIe lane is also treated as a memory bus in modern computer systems, but it transfers 4KB or larger data packets between the CPU and the SSD for I/O transactions. 
Since the granularity of I/O accesses is a page or a block, user applications can only access the underlying SSD by going through the entire storage stack of the OS, 
which includes an I/O runtime library, file system, and block layer, atop an NVMe driver. 
The NVMe driver manages the transfers of data packets over PCIe, and communicates with the NVMe controller in the SSD through the PCIe \emph{baseline address registers} (BARs), including 
doorbell registers, queue attributes, target addresses for each queue, and the NVMe controller information \cite{ellefson2013nvm}. The internal hardware details of ULL-Flash will be explained in Section \ref{sec:ull-hardware}. 

\begin{figure}
	\centering
	\includegraphics[width=1\linewidth]{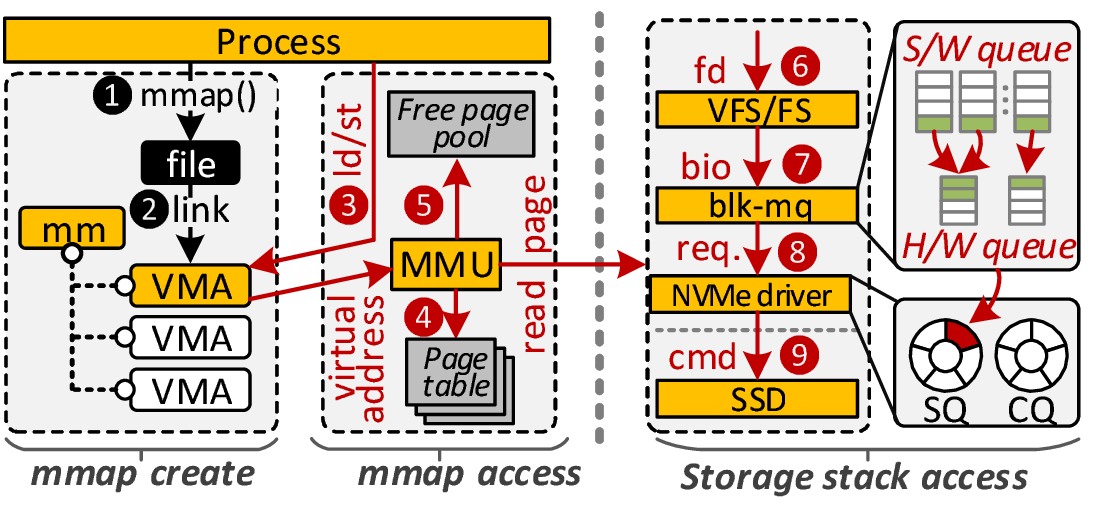}
	\vspace{-20pt}
	\caption{\label{fig:bck_mmap}Software support.\vspace{-15pt}}
	\vspace{-5pt}
\end{figure}

\begin{figure*}
\centering
\subfloat[SSD internal parallelism.]{\label{fig:bck_parallel}\rotatebox{0}{\includegraphics[width=0.32\linewidth]{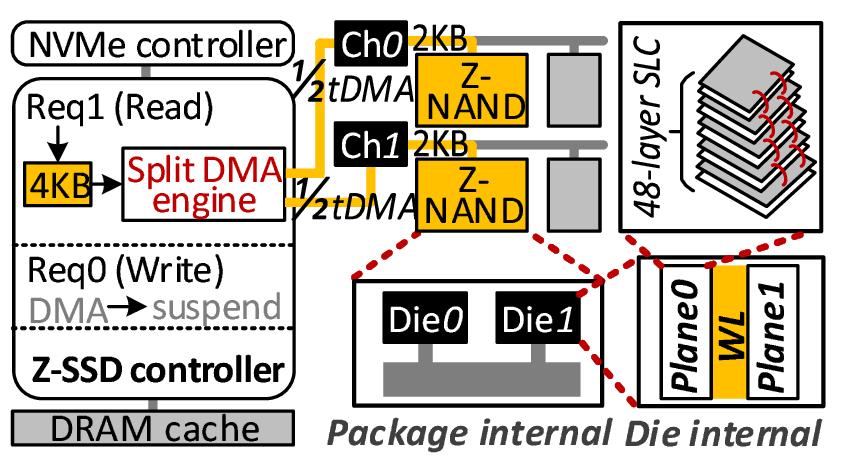}}}
\subfloat[NVMe queue and protocol management.]{\label{fig:bck_nvme}\rotatebox{0}{\includegraphics[width=0.48\linewidth]{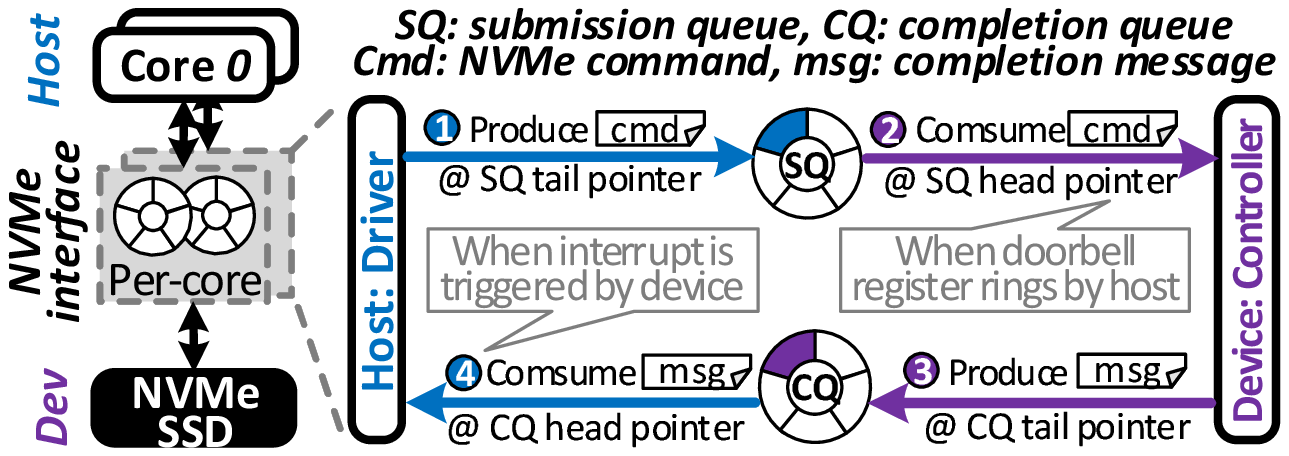}}}
\hspace{1pt}
\subfloat[Flash firmware.]{\label{fig:bck_firmware}\rotatebox{0}{\includegraphics[width=0.18\linewidth]{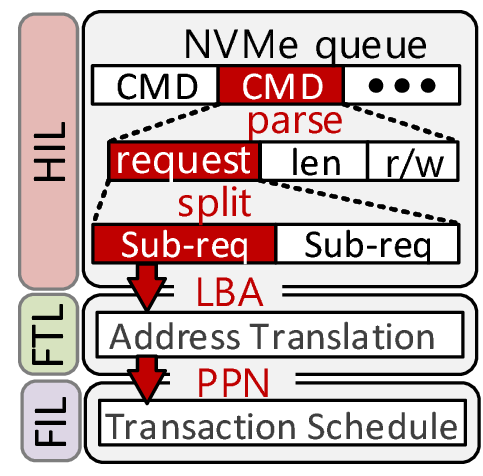}}}
\caption{\label{fig:bck_ssd}Overview of ULL-Flash and NVMe datapath.}
\vspace{-20pt}
\end{figure*}

\subsection{Support for Persistent Memory Expansion}
Figure \ref{fig:bck_mmap} illustrates the software support and storage stack that user applications require for expanding NVDIMM with SSD. 
The \emph{memory-mapped file} (MMF) module in Linux, also referred to as \texttt{mmap}, can be used to expand the persistent memory space of NVDIMM with SSD. 
If a process calls \texttt{mmap} with a file descriptor (\texttt{fd}) for SSD (\whitecircled{1}), the MMF creates a new mapping in its process address space, 
represented by a memory management structure (\texttt{mm\_struct}), by allocating a virtual memory area (VMA) to the structure (\whitecircled{2}). 
In other words, the MMF links \texttt{fd} to the VMA, by establishing a mapping between the process memory and the target file. 
When the process accesses the memory designated by the VMA (\redcircled{3}\redcircled{4}\redcircled{5}), this triggers a page fault (if the data is not available in NVDIMM). 

When a page fault occurs, the page fault handler is invoked and allocates a new page to the VMA. 
Since the VMA is linked to the target file, the page fault handler retrieves the file metadata (\texttt{inode}) associated with \texttt{fd} and acquires a lock for its access (\redcircled{6}). 
The MMU interacts with a fault handler of the file system to read a page from the SSD. 
The file system initializes a block I/O request structure, called \texttt{bio}, and submits it to the multi-queue block I/O queueing (\emph{blk-mq}) layer, 
which schedules I/O requests over multiple software queues (\redcircled{7}). 
Depending on the design of the target system, one or more software queues can be mapped to a hardware dispatch queue (\redcircled{8}), managed by the NVM controller that exists within the SSD (\redcircled{9}). 
Once the service of the I/O request (i.e., \texttt{bio}) is completed, and the actual data is loaded into a new region of the allocated page memory, the page fault handler creates a page table entry (PTE), records the new page address in the PTE, and resumes the process.

The MMF can be used to expand the persistent memory space of NVDIMM with SSDs. 
However, this approach can potentially negate most of the benefits that would be brought by ULL-Flash, due to the high overheads caused by page faults, file systems, context switching, and data copies.

\subsection{ULL-Flash}
\label{sec:ull-hardware}
\noindent \textbf{Hardware details.} 
All state-of-the-art SSDs typically employ a large number of flash packages and connect them to multiple system buses, referred to as \emph{channels}. 
Each flash package contains multiple dies and planes for fast response time and low latency, as illustrated in Figure \ref{fig:bck_parallel}. 
To deliver massive parallelism and hence high I/O performance, an SSD spreads a given set of I/O requests from the host across multiple channels, packages, dies, and even planes.

ULL-Flash also adopts this multi-channel and multi-way architecture, but it optimizes the datapath and channel stripping \cite{cheong2018flash}. 
More specifically, ULL-Flash splits a 4KB I/O request from the host into two operations and issues them to two channels simultaneously; doing so can effectively reduce the DMA latency by half. 
In addition, while most high-performance SSDs employ multiple-level cell (MLC) or triple-level cell (TLC), ULL-Flash employs a new type of flash medium, called \textit{Z-NAND} \cite{cheong2018flash}. 
Z-NAND leverages a 3D-flash structure to provide a single-level cell (SLC) technology and optimizes the I/O circuitry and memory interface to enable short latency. 
Specifically, Z-NAND uses 48 stacked word-line layers, referred to as the vertical NAND (V-NAND) architecture, to incarnate an SLC memory. 
Thanks to its unique flash architecture and advanced fabrication technology, 
the read and write latencies of Z-NAND (i.e., 3$\mu$$s$ and 100$\mu$$s$) are 15$\times$ and 7$\times$ lower than the V-NAND flash memory, respectively \cite{samsung2017znand}. 

ULL-Flash employs large DRAM in front of its multiple channels and exposes its internal parallelism, low latency, and high bandwidth (through the NVMe interface), which are managed by multiple interface controllers and firmware modules. 
Note that the DRAM management is tightly coupled with handling the NVMe protocol. Based on the definition of NVMe, the same data can be in both the host-side DRAM and the SSD-internal DRAM after the underlying ULL-Flash controller or firmware performs DMA for data transfer.


\noindent \textbf{I/O connection to CPU.} 
Figure \ref{fig:bck_nvme} illustrates the per-core NVMe queue and communication protocol. 
An NVMe queue consists of a pair of \emph{submission queue} (SQ) and \emph{completion queue} (CQ), each with 64K entries \cite{huffman2013nvm}. 
These are simple FIFO queues, and each entry is referenced by a \emph{physical region page} (PRP) pointer \cite{ellefson2013nvm}. 
If the request size is larger than a 4KB NVMe packet, the data can be referenced by a list of PRP pointers instead of a single PRP pointer. 
When a request arrives at the SQ, the host 
increments its tail (pointer) and rings the corresponding doorbell of ULL-Flash, 
so that the NVMe controller can synchronize the storage-side SQ, which is logically paired with the host-side SQ. 
Since the data for each entry exist in the host-side DRAM (pointed by a PRP pointer), the ULL-Flash handles the DMA for the I/O request, and then the underlying Z-NAND and firmware serve the request. 
Once the service is completed, the NVMe controller moves the tail of the CQ (paired with the SQ) and informs the host of the event over a message signaled interrupt (MSI). 
The host then jumps to an interrupt service routine (ISR) and synchronizes the CQ tail. 
The ISR completes the request, advances the head of the CQ (and releases the buffer data), and rings the doorbell to notify the ULL-Flash of the completion of the host-side I/O processing. 
Finally, the NVMe controller releases the internal data and advances the head pointer of the CQ. The NVMe interface has no knowledge of the data cached in the host-side DRAM, while the data for each I/O request can reside in the host-side DRAM.
Therefore, even if I/O requests can be served by the host-side DRAM, the NVMe interface 
obliviously
enqueues the requests and processes them.

\begin{figure}
	\centering
	\includegraphics[width=1\linewidth]{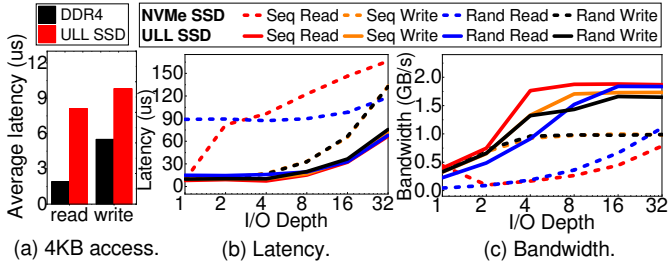}	
	\vspace{-10pt}
	\caption{\label{fig:zssd_overall}ULL-Flash versus NVMe SSD.\vspace{-5pt}}
	\vspace{-15pt}
\end{figure}

\noindent \textbf{Firmware.} Figure \ref{fig:bck_firmware} shows a general 
firmware architecture implemented in ULL-Flash. 
At the top of the firmware layers within the ULL-Flash, the \emph{host interface layer} (HIL) is responsible for parsing the NVMe commands and managing the queues by collaborating with the internal NVMe controller \cite{jung2018simplessd}. 
This layer also splits an I/O request, which can be of 
any length, into sub-requests. 
The size of a sub-request matches the unit I/O size that the underlying firmware module manages. 
The parsed 
separate 
requests are forwarded to the \emph{flash translation layer} (FTL) \cite{kim2002space}. 
The FTL translates a given \emph{logical block address} (LBA) to a \emph{physical page number} (PPN). 
After translating the address of each sub-request into a PPN, the flash interface layer (FIL) submits the request and manages its transactions, which constitutes multiple flash commands
such as row/column addresses, I/O commands, administrative commands, and DMA transfers. 
During this I/O processing, FTL/FIL can stripe the requests across multiple internal resources (\textit{e.g.}, channels, packages, dies, planes, etc.), achieving both low latency and high bandwidth.

\section{Motivation and Challenges}
\label{sec:motiv}

In this section, we explain 
why ULL-Flash can be used for a large working memory solution, and discuss what 
challenges 
the conventional software-assisted solutions face to expand the persistent memory by 
integrating NVDIMM with ULL-Flash.

\subsection{ULL-Flash Performance Characterization}

We evaluated 
a real 800 GB Z-SSD \emph{prototype} (\cite{ULLSSD} as ULL-Flash) and analyzed its performance characteristics.
We then compared the performance characteristics of ULL-Flash with those of a high performance NVMe SSD (Intel NVMe 750 \cite{intel750series}) using a Flexible I/O Tester \cite{axboe2017flexible}. 
Both 
the devices use four PCIe3.0 lanes (1GB/lane) and are evaluated by a system that has a single 4GHz CPU \cite{intel4790cpu}. 
The collected performance characteristics are plotted in Figure \ref{fig:zssd_overall}, under the sequential and random read/write accesses.
We also evaluated the performance with varying I/O queue depths (1$\sim$32). 
The request sizes equal to that of the NVMe packet payload (4KB).

As shown in Figure \ref{fig:zssd_overall}a, we observe that ULL-Flash exhibits 8 $\mu$s and 10 $\mu$s for 4KB read and write latencies with 1$\sim$4 queue depth at the user-level.
That is, such read and write latencies of ULL-Flash are only 3.3$\times$ and 79\% longer than the real read/write latencies (4KB-sized) of single DDR4-2133 DIMM \cite{DDR4-2133} on the same testbed. This significant latency advantage makes ULL-Flash a promising replacement for conventional SSD to expand the persistent memory space of NVDIMM with storage. As shown in Figure \ref{fig:zssd_overall}b, ULL-Flash maintains such latency characteristics under different I/O depths in a predictable and sustainable manner, while NVMe SSD experiences significantly increased latencies as the I/O depth increases (up to 155 $\mu$s). Figure \ref{fig:zssd_overall}c compares the bandwidth trends of ULL-Flash with those of NVMe SSD. 
For read and write accesses, ULL-Flash offers 115\% and 137\% higher average bandwidth than NVMe SSD.
These plots also indicate that ULL-Flash reaches its peak bandwidth with only a few NVMe commands, whereas an NVMe SSD does not achieve such peak bandwidth for random read accesses, even if we increase the queue depth to 32 entries. We also observe that \emph{the number of requests in ULL-Flash waiting queue is 4 for most accesses (cf. Figure \ref{fig:zssd_overall}c)}.
ULL-Flash can support only 16 outstanding requests for random read accesses. 
We believe that this characteristic can make the NVMe queue management simple and amenable to be implemented in hardware. 


\subsection{Software-Based Memory Expansion}
\label{sec:sw-mem-expansion}
To evaluate the performance of an existing software-based memory expansion, 
we configure an MMF-based host system with the real devices. Our evaluation system integrates three SSDs (including a SATA SSD \cite{intel535series}, in addition to ULL-Flash and NVMe SSD) and employs two 8GB DRAM ranks, each with eight banks operating at 1.6GHz. 
The ULL-Flash is used to expand memory space over \texttt{mmap}.

\noindent \textbf{Benchmarks.} We use \texttt{mmap-benchmark}, which is designed to evaluate the performance of \texttt{mmap} with a set of microbenchmarks \cite{matt2014mmap}. 
Each of \texttt{seqRd} and \texttt{seqWr} creates a single thread, 
and then performs sequential read and write operations.
In contrast, each of \texttt{rndRd} and \texttt{rndWr} creates four threads, each simultaneously
performing random read and write operations.
We also tested \texttt{SQLite-benchmark}, which is a benchmark 
for a widely-used DBMS (\texttt{SQLite}) \cite{sqlite2017}. 
The workload details will be explained in Section \ref{sec:experiment}.

\begin{figure}
\centering
\vspace{-10pt}
\subfloat[mmap-bench.]{\label{fig:mmap_test}\rotatebox{0}{\includegraphics[width=0.48\linewidth]{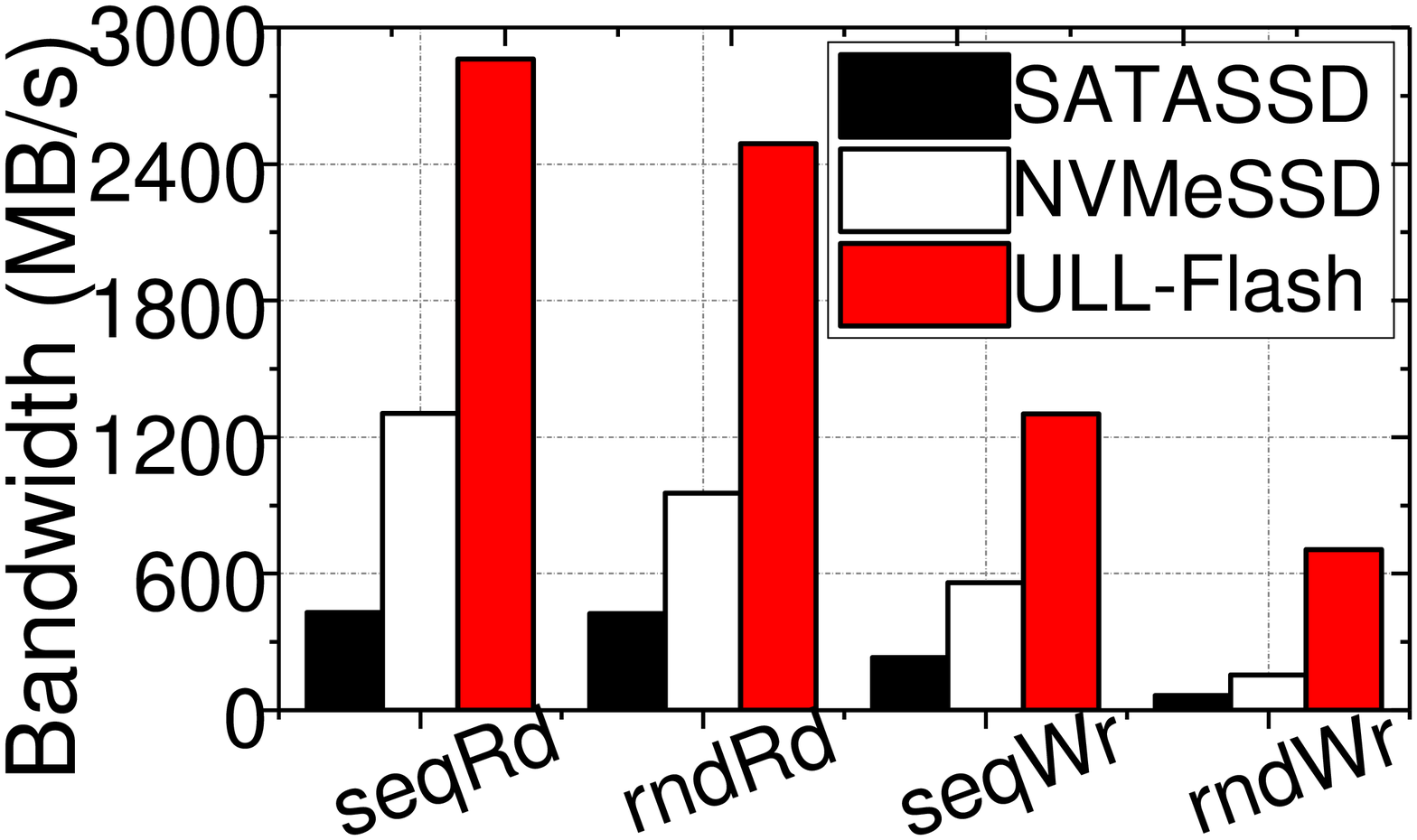}}}
\hspace{2pt}
\subfloat[SQLite.]{\label{fig:sqlite}\rotatebox{0}{\includegraphics[width=0.48\linewidth]{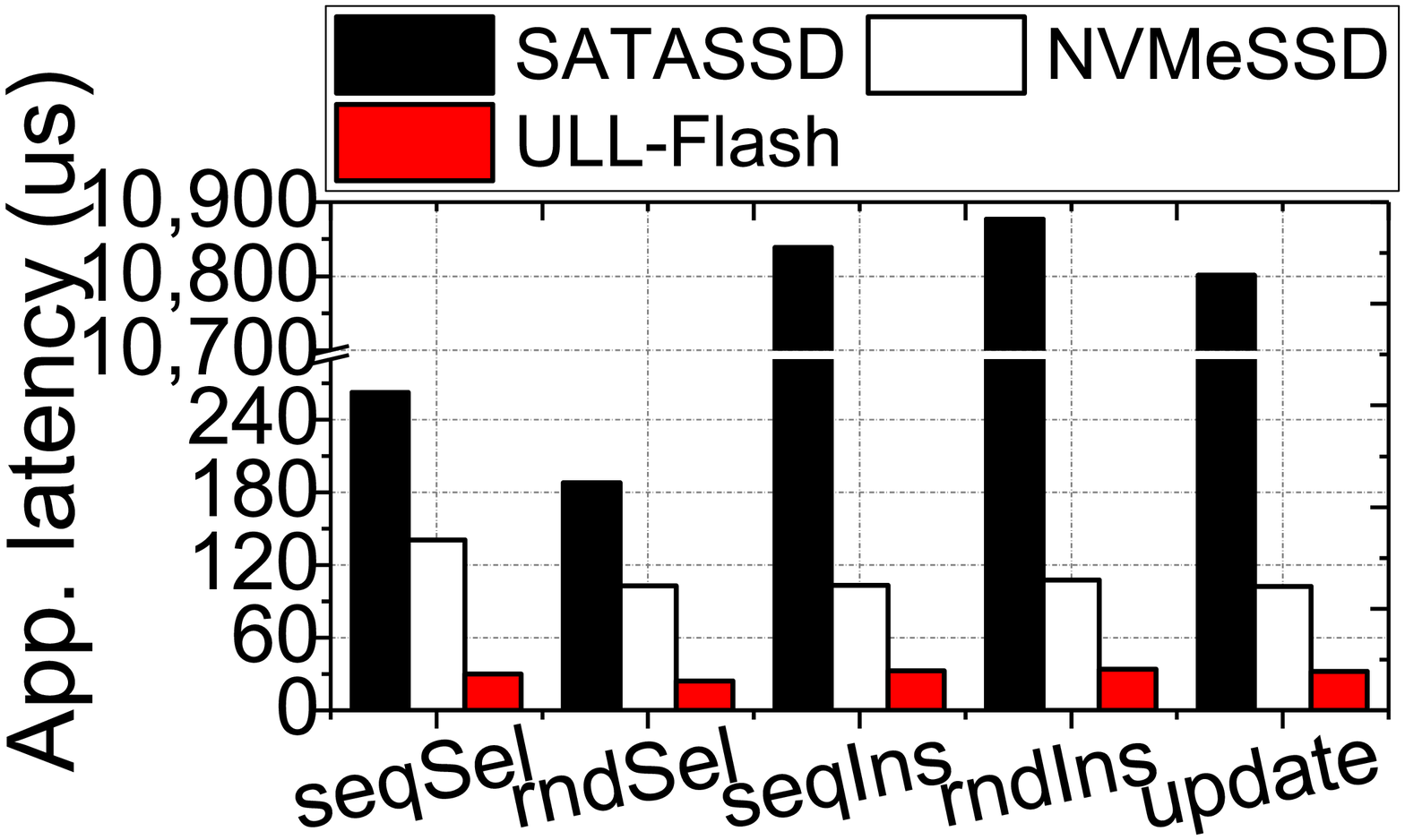}}}
\caption{\label{fig:real_device_comparison} MMF-based system performance. \vspace{-5pt}}
\vspace{-15pt}
\end{figure}

\noindent \textbf{Performance.} 
Figures \ref{fig:mmap_test} and \ref{fig:sqlite} show the performances of \texttt{mmap-benchmark} (bandwidth) and \texttt{SQLite} (transaction latency), respectively. 
As shown in Figure \ref{fig:mmap_test}, ULL-Flash exhibits 399\% and 118\% higher bandwidth than SATA and NVMe SSDs, respectively, in the MMF-system. 
\texttt{seqRd} and \texttt{seqWr} exhibit the performance near the peak bandwidth of the SSDs \cite{ULLSSD,intel750series,intel535series}, but 
they significantly degrade performance while executing \texttt{rndRd} and \texttt{rndWr}. 
This is because \texttt{seqRd} and \texttt{seqWr} pull the data in a sequential manner to the file system's buffer cache, 
and this helps us hide the performance degradation of SSDs for byte-based I/O accesses. 
In addition, the average I/O queue depths of \texttt{mmap-benchmark} and \texttt{SQLite} are one and four, respectively, 
which can better leverage the benefits of read ahead. 
Similarly, Figure \ref{fig:sqlite} shows that the average latency of ULL-Flash for \texttt{SQLite} (per transaction) is lower than those of SATA and NVMe SSDs by 95\% and 72\%, respectively. 

\begin{figure}
\centering
\vspace{-5pt}
\subfloat[Software overheads.]{\label{fig:oval_motiv1}\rotatebox{0}{\includegraphics[width=0.48\linewidth]{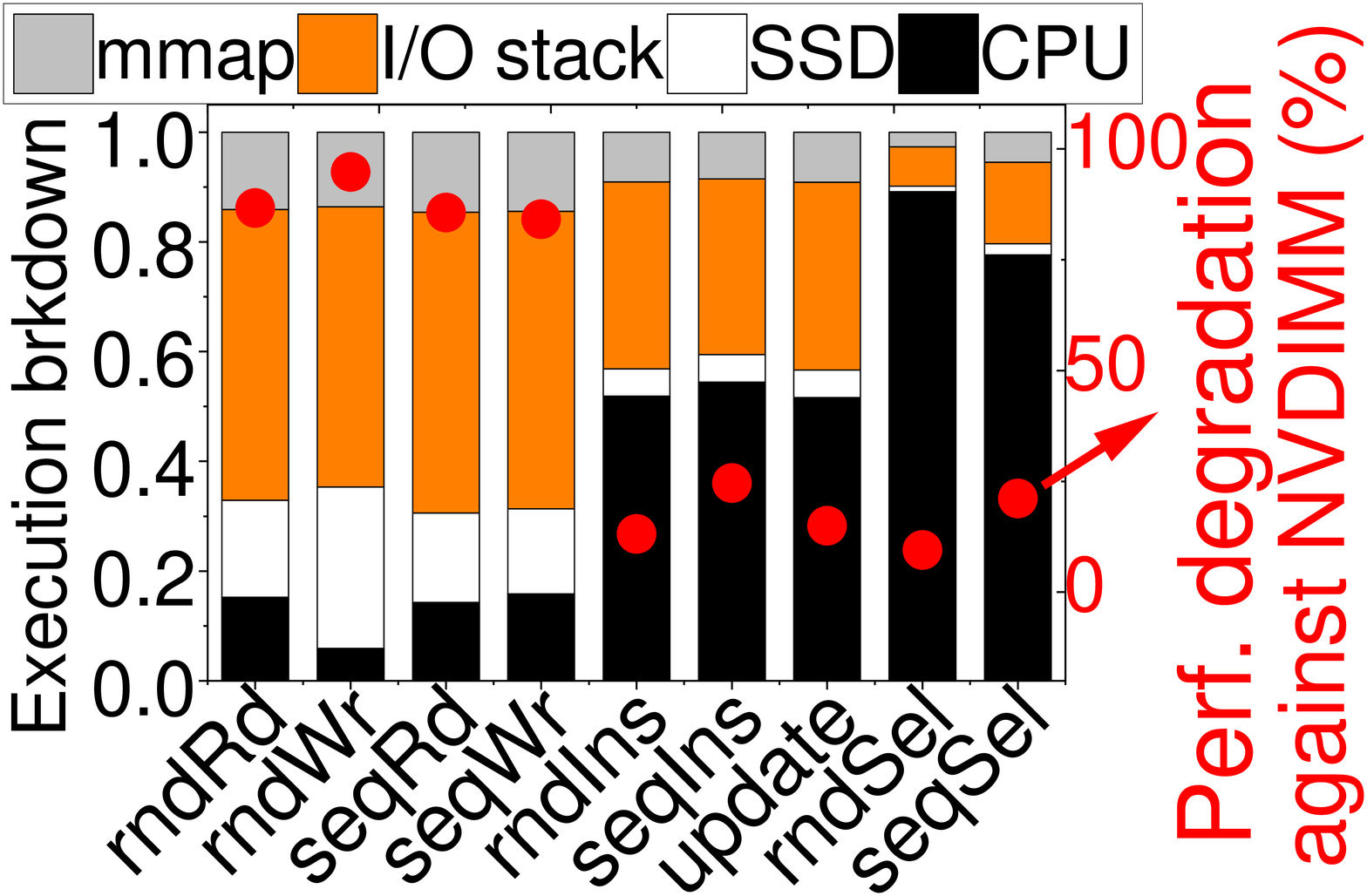}}}
\hspace{1.5pt}
\subfloat[IPC of bypassing.]{\label{fig:oval_motiv2}\rotatebox{0}{\includegraphics[width=0.49\linewidth]{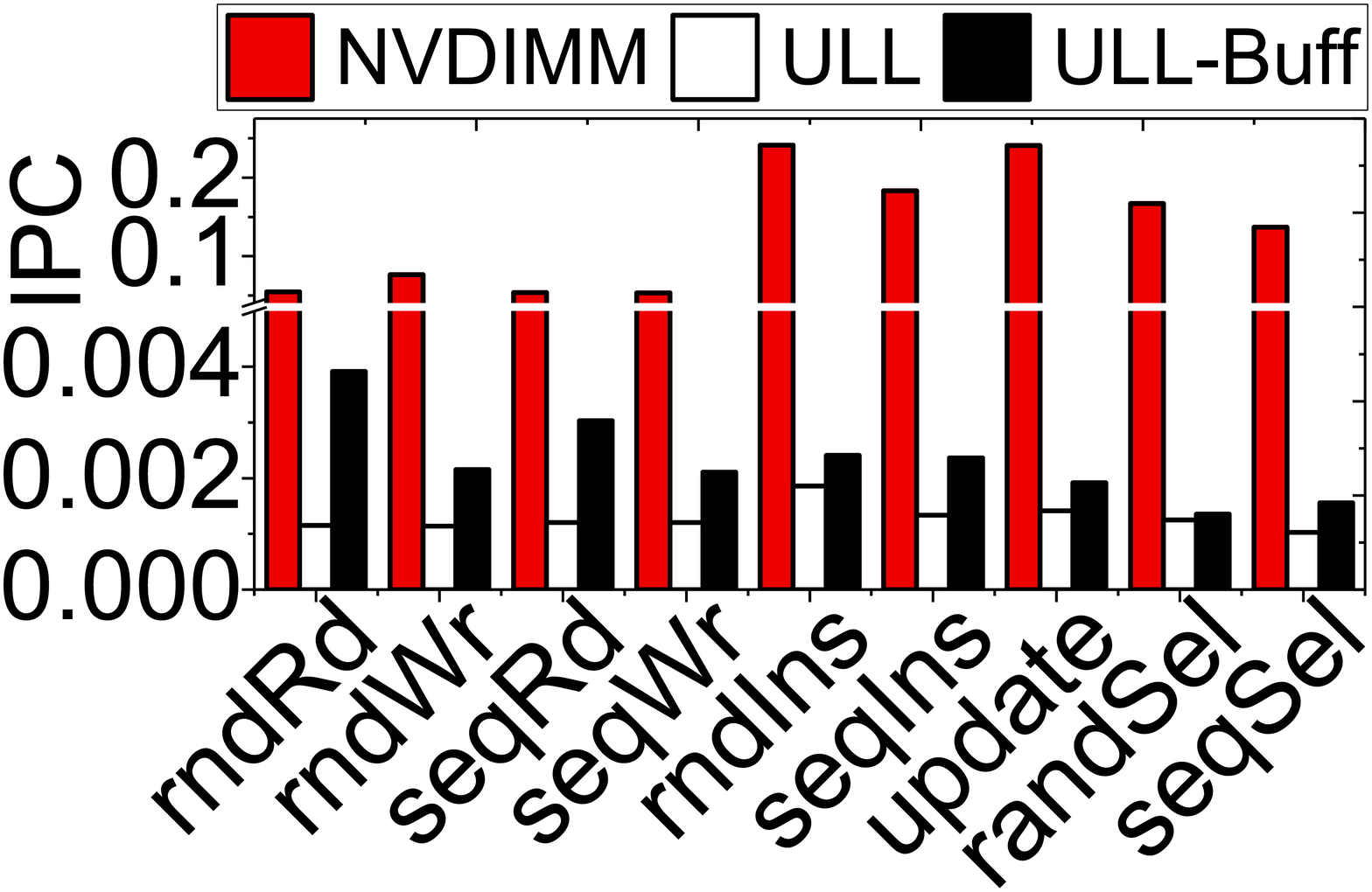}}}
\caption{\label{fig:over_motiv}Challenges of using ULL-Flash.\vspace{-5pt}}
\vspace{-15pt}
\end{figure}

\noindent \textbf{Analysis on ULL-Flash overhead.} 
\newedit{Figure \ref{fig:oval_motiv1} further decomposes the total execution time of user applications into a mmap processing time (i.e., context switch and page fault handling), an I/O stack time (i.e., filesystem, blk-mq layer, and NVMe driver), a ULL-Flash access time, and an application computation time. }
\newedit{For better understanding, the figure also analyzes how much the ULL-Flash-based MMF system degrades the overall performance, compared to the NVDIMM-based system.} 
Since \texttt{rndSel} and \texttt{seqSel} spend most of their execution on the DBMS-side computation, their CPU cycles account for 83\% of the total execution time, on average. 
However, the remaining workloads in \texttt{mmap-benchmark} and \texttt{SQLite} take 13\% and 53\% of the execution, respectively, 
while ULL-Flash only accounts for 13\% of the total latency, on average. 
\newedit{Note that the system overhead imposed by MMF (\texttt{mmap} and I/O stack) accounts for 69\% of the total execution time.
This is because MMF is involved in many software operations including multiple page fault handling, context switches, address translations (i.e., page table, filesystem and FTL), boundary checks, and permission checks \cite{huang2015unified}. The context switches are one of the main contributors to increase I/O latency \cite{le2017latency}. On the other hand, the queuing mechanism and NVMe communication protocol in I/O stack are optimized for throughput rather than I/O latency \cite{zhang2018flashshare}. The software operations of MMF consume 15$\sim$20 us \cite{huang2015unified}, which is around 6$\times$ longer than ZNAND access latency (3 us). 
}


\subsection{Software Overhead}
To remove the software overheads brought by MMF, we can \textit{bypass} the entire storage stack and simulate the underlying ULL as a memory module to directly serve the load/store instructions.
Figure \ref{fig:oval_motiv2} shows the CPU-side performance for three different bypass strategies: 
(1) NVDIMM only (\texttt{NVDIMM}), 
(2) ULL-Flash (\texttt{ULL}), and 
(3) ULL-Flash with a page buffer, which is essentially a small DRAM (\texttt{ULL-buff}).
For this evaluation, we use the same workloads used in Section \ref{sec:sw-mem-expansion}. 
The results collected with \texttt{mmap-benchmark} indicate that the average instructions per cycle (IPC) values for the \texttt{ULL}- and \texttt{ULL-buff}-based systems are only 0.001 and 0.003, respectively, while the \texttt{NVDIMM}-based system offers an average IPC of 0.06 (i.e., 98\% and 95\% degradation).

When evaluating \texttt{SQLite}, we observed that \texttt{ULL} and \texttt{ULL-buff} degrade IPC compared to \texttt{DRAM} by 140$\times$ and 101$\times$, respectively. 
The load and store instructions take 51\% of the total number of executed instructions, and all these load/store instructions for the workloads we tested are due to the relatively long ULL-Flash operations. 
Note that the Z-NAND latency (3$\mu$$s$) is much shorter than that of conventional flash, but it is 3.3$\times$ longer than the latency of NVDIMM for 4KB access. 
While a page cache can potentially hide the page access delay, we observe that a large fraction of the load/store instructions suffer from the page cache misses, due to the poor data locality exhibited by \texttt{mmap-benchmark} and \texttt{SQLite}.

The goal of HAMS is to (1) remove the \texttt{mmap} and storage stack overheads from the MMF-system and (2) reduce the number of stalled instructions by caching the memory references in NVDIMM directly and by automating the mapping between ULL-Flash and NVDIMM.

\section{Memory Over Storage}
\label{sec:highlevelview}

\ignore{
$$$$$$$$$$$$$$$$$$$$$$$$$$$$$$$$$$$ benchmark
Dear Professor Jung,

I use two benchmarks:
microbenchmark comes from mmap-benchmark https://github.com/exabytes18/mmap-benchmark. The source codes are written for testing mmap function. I modify the source codes to enable in-memory operation.
The data volume that mmap-benchmark accesses is 512MB. 4KB is the minimum access granularity. seq-rd and seq-wr create single thread and access data in sequential. rnd-rd and rnd-wr create 4 threads (in single core) and make each thread to access the data independently and randomly.

SQLite benchmark comes from LevelDB benchmark http://www.lmdb.tech/bench/microbench/benchmark.html. SQLite provides the primitives to use mmap or in-memory operation. I modify the sqlite3_open() function with ":memory:" to enable in-memory option. In addition, I use "PRAGMA mmap_size = " to enable mmap option.
SQLite creates database with 5 million pairs of key-values. each value size is 100B. For performance optimization, SQLite sets its locking mode to exclusive and also enable SQLite's write-ahead logging.
$$$$$$$$$$$$$$$$$$$$$$$$$$$$$$$$$$$ benchmark
}

HAMS is aimed to automate all necessary hardware for the expansion of the persistent memory by integrating NVDIMM and ULL-Flash, while reducing the energy consumption and maintaining the consistency on the resulting heterogeneous memory space. In this section, we give an overview of the baseline design and aggressive integration of HAMS. 


\subsection{HAMS Overview}
\begin{figure}
\centering
\includegraphics[width=0.95\linewidth]{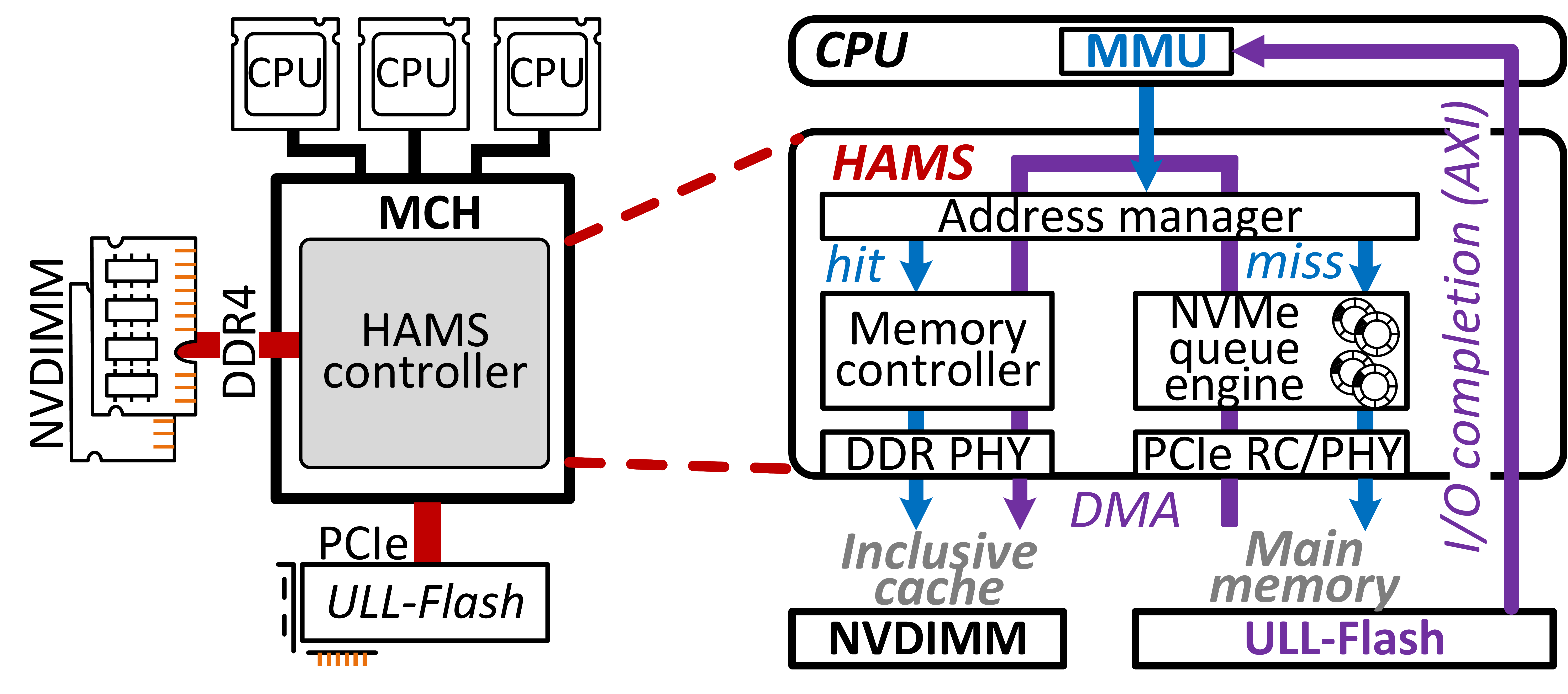}
\vspace{-5pt}
\caption{\label{fig:over_LX}Overview of baseline HAMS.\vspace{-15pt}}
\vspace{-5pt}
\end{figure}

Figure \ref{fig:over_LX} shows the baseline architecture of HAMS. 
\newedit{HAMS resides in MCH, which implements an address manager, an NVDIMM memory controller and PCIe root complex. 
The address manager offers a 64-bit byte-addressable address space by exposing the storage capacity of ULL-Flash to MMU. 
It also utilizes a memory space of NVDIMM as an inclusive cache for ULL-Flash with an integrated tag-array. 
To implement MoS, the address manager employs a simple hardware cache logic that coordinates NVDIMM and ULL-Flash to serve incoming memory requests. 
Note that a memory request can be generated by either MMU or ULL-Flash, and thus, they should be processed differently.
When the NVMe controller (in ULL-Flash) generates a memory request, the NVMe controller extracts the NVDIMM address of the target data by referring to PRP(s) that the address manager handles and records it in the request. HAMS then directly forwards the request to access NVDIMM based on the recorded NVDIMM address.
On the other hand, HAMS checks the memory address of MMU's request by examining its MoS tag-array. 
If the requested memory address hits in the MoS tag-array, the request is directly served by the data from NVDIMM. 
Otherwise, HAMS secures an NVDIMM space to accommodate the incoming request by evicting data to ULL-Flash. HAMS also fetches target data from ULL-Flash to NVDIMM for read requests. 
Once the data transfer from ULL-Flash to NVIDMM (or vice versa) is completed, HAMS informs MMU of the completion 
so that MMU can retry the stalled instruction.
}


\subsection{ULL-Flash Archive Management}
\label{subsec:manage}
The power failure management for persistency control is central to a key design of HAMS. 
While NVDIMM's data is stored and restored by its on-board NVM controller, 
NVMe storage employs a different mechanism to handle power failures. 
Specifically, the data persistency and I/O atomicity of an SSD are guaranteed by a file system. 
The file system and other kernel related components in typical support persistency using journaling \cite{woodhouse2001jffs}. 
Since HAMS removes the MMF and file system support, the data in the SSD-internal DRAM can be lost upon a power failure. 
While HAMS can enforce data persistency by tagging force unit access (FUA) per request \cite{stevens2005information}, 
doing so can significantly degrade the ULL-Flash performance. \newedit{This is because FUA enforces serializing all outstanding requests to be written to the underlying flash media.}
Another issue in the design of HAMS is related to hardware implementations of the NVMe protocol. 
Since NVMe data structures, including SQs, CQs, and BARs, are mapped to a memory region of NVDIMM, 
all data and queue information can be unintentionally overwritten by any of user applications or the OS. 
This can make the hardware-based NVMe management in HAMS vulnerable. 
In addition, the data in NVDIMM can be inconsistent if HAMS and ULL-Flash simultaneously access the same page frame. 

\begin{figure}
\centering
\includegraphics[width=1\linewidth]{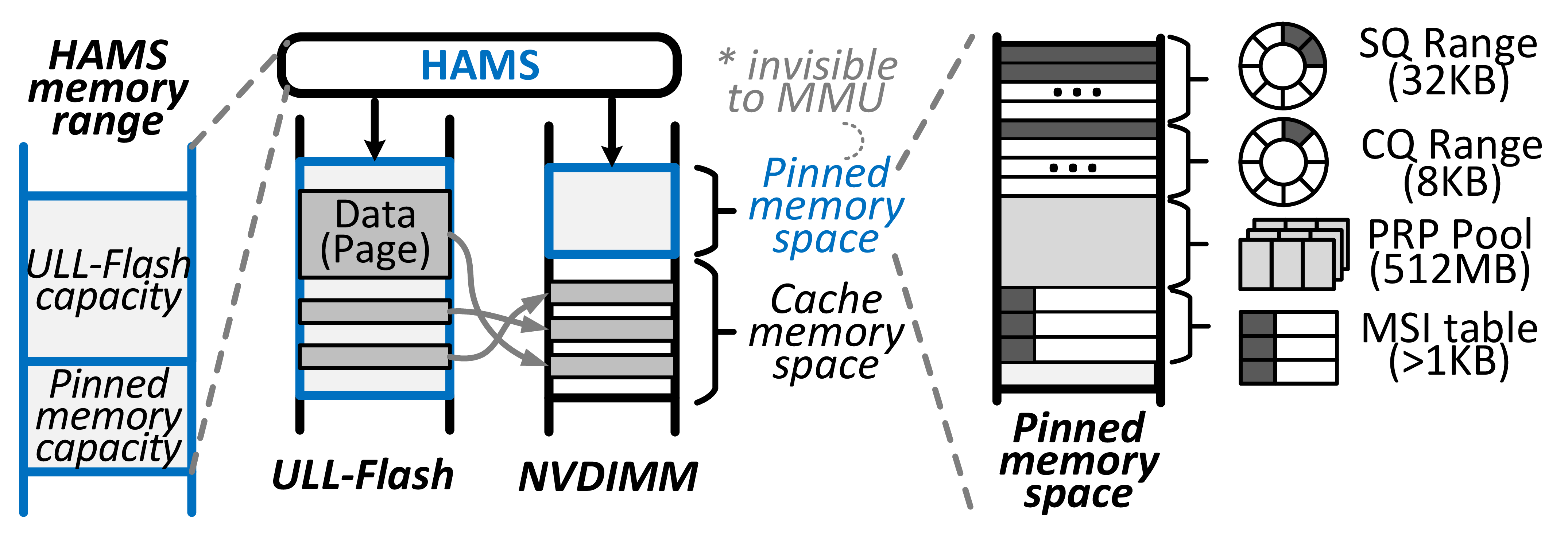}
\vspace{-20pt}
\caption{\label{fig:impl_init1}Address management.\vspace{-15pt}}
\vspace{-5pt}
\end{figure}

\newedit{
We propose a set of designs to address the aforementioned challenges. Specifically, to protect data against power failure, we integrate super-capacitors in ULL-Flash to flush data in the volatile DRAM buffer to the persistent flash media. We also utilize the NVMe data structure to recover the I/O requests, which are corrupted by power loss.
On the other hand, to resolve the vulnerability issue of the NVMe data structure, we pin a specific memory region of NVDIMM to store the NVMe data structure and make it invisible to the MMU. }
As shown in Figure \ref{fig:impl_init1}, this pinned memory region includes the ring buffers for the SQs and CQs, PRP pool and MSI table, 
and it is mapped to the upper memory part of NVDIMM (around 512MB in our design). 
On the other hand, the remaining NVDIMM memory area is mapped to the MoS address space by HAMS. 
During the initialization process, HAMS reviews the pinned memory region, in particular the SQ and CQ buffers including the head and tail pointers. 
If there is no power failure, the SQ and CQ tail pointers should refer to the same offset of their queue entries to avoid a violation of NVMe queue management and consistency \cite{huffman2012nvm}. 
However, if a power failure occurs, HAMS is able to detect all pending requests by checking the offset differences between the SQ/CQ tail pointers in the MMU-invisible space of NVDIMM (cf. Figure \ref{fig:impl_init1}). During the power restoration, HAMS needs to reissue all pending requests to the underlying ULL-Flash for data persistency and consistency. 
To protect the memory to which the data is being transferred, HAMS keeps track of the DMA status by configuring a bit per each entry of the MoS tag-array, 
which is referred to as \emph{busy bit}. 
This bit is set to 1 whenever the NVMe engine issues a command, and it can be cleared when HAMS updates the CQ's head pointer. 
Thus, if the busy bit is set, HAMS will exclude the corresponding page from being evicted. This guarantees that the data is consistent when ULL-Flash accesses the page frame via PRP.

\ignore{
I think duplicating data over PRP region has two benefits:
1. protect memory on DMA;
2. as PRP region can serve DMA, the NVDIMM cache lines can be available earlier to serve other memory accesses.
}


\subsection{Aggressive Integration of HAMS}
The baseline design of HAMS explained so far includes a hardware automation of cache logic in the MCH by leveraging the conventional DDR and PCIe controllers, thus offering a large working memory space. 
While this design strategy does not require any modification to the existing storage and memory devices, it brings two inefficiencies from an architectural perspective: 
(1) the overheads imposed by data transfer and 
(2) the energy inefficiency brought by the SSD-internal DRAM. 
First, in case of a cache miss, the target data needs to go through the DDR4 module (e.g., the memory controller and DDR4 PHY) and the PCIe module (root complex, transaction layer, data link layer and physical layer). 
While the peak bandwidth of DDR4 \cite{standard2012jesd79} is 20 GB/s per channel, 
ULL-Flash (including most NVMe SSD products) uses PCIe 3.0 with 4 lanes, which makes the maximum bandwidth of NVMe 4 GB/s. 
Thus, in case of a cache miss, the performance of HAMS can be capped by the peak PCIe bandwidth. 
In addition, the raw data of NVDIMM should be encoded and encapsulated into a PCIe packet, which also makes the HAMS latency longer in case of a cache miss.

\begin{figure}
\centering
\vspace{-5pt}
\subfloat[DMA overhead.]{\label{fig:over_TX_chal}\rotatebox{0}{\includegraphics[width=0.33\linewidth]{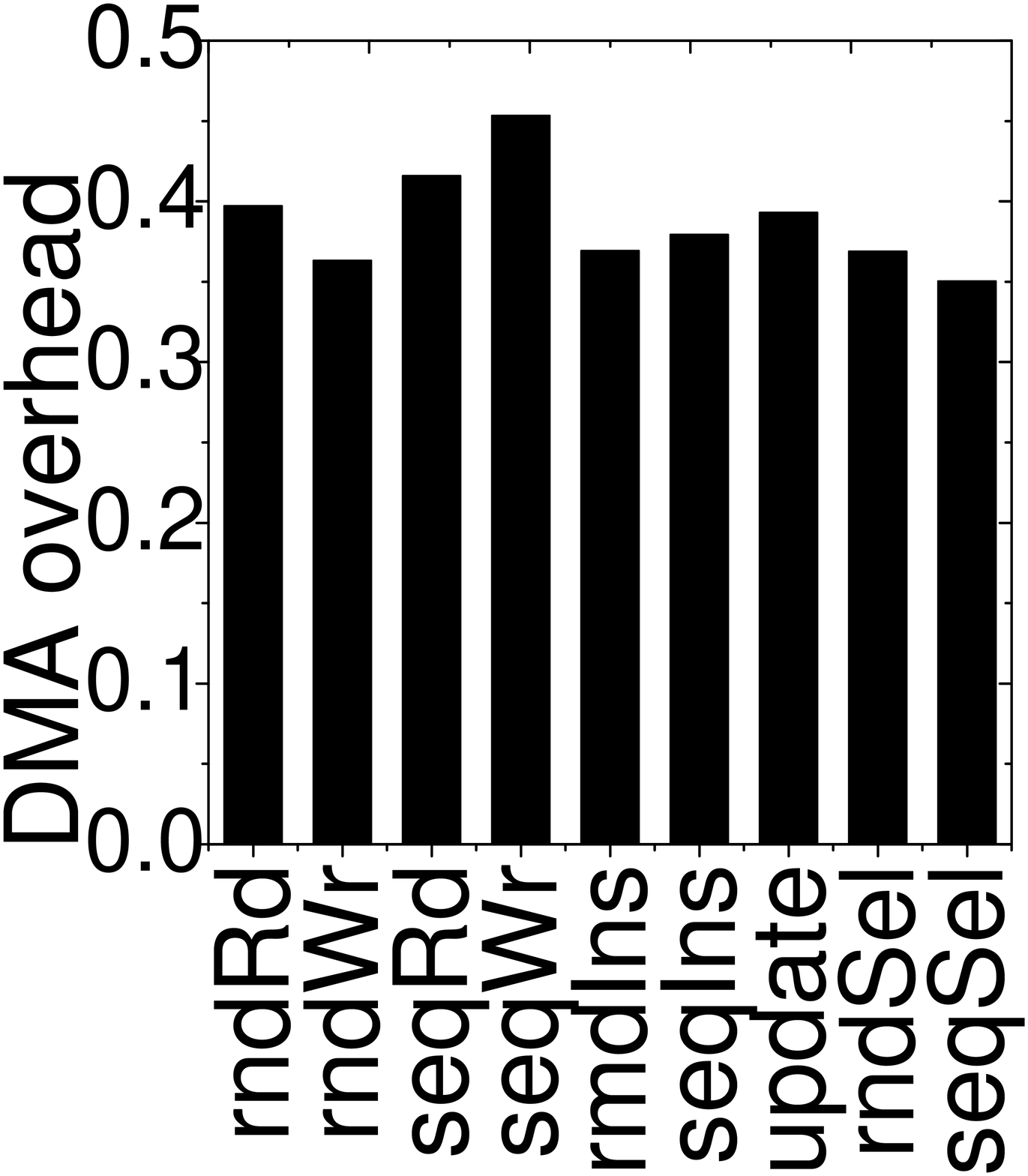}}}
\hspace{1pt}
\subfloat[Aggressive integration.]{\label{fig:over_TX}\rotatebox{0}{\includegraphics[width=0.65\linewidth]{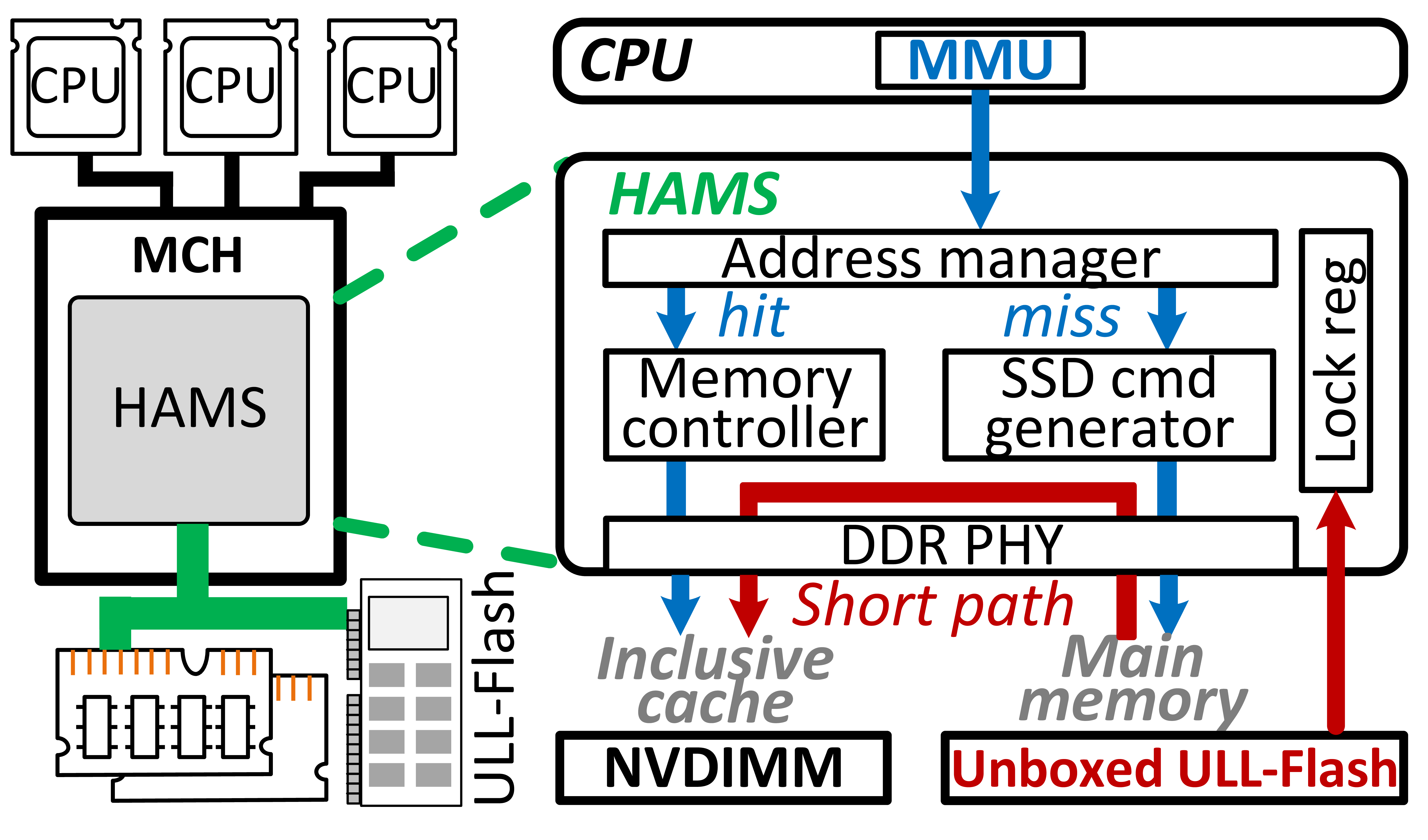}}}
\caption{\label{fig:over_TX_sum}Challenges and aggressive integration.\vspace{-10pt}}
\vspace{-10pt}
\end{figure}

Figure \ref{fig:over_TX_chal} shows the fraction of data movement latency in the average memory access time (AMAT) under the execution of the workloads selected from Section \ref{sec:motiv}. 
It can be observed that the interface latency taken by moving data between the NVMe controller and the DDR4 controller constitutes 39\% of the total AMAT, 
which can degrade the HAMS performance. 
Another drawback of the baseline design of HAMS is that, even if HAMS already holds the data in NVDIMM, the data will still be copied to the SSD-internal DRAM. 
While this would improve performance under the block storage use-case (with a file system), it would also introduce extra energy consumption and increase the internal complexity of the SSD. 
Note that the SSD-internal DRAM requires 17\% more power than a flash complex consisting of 32 flash chips.

To address these challenges, we propose to remove the SSD-internal DRAM that is used for data buffering, introduce a new register-based interface (instead of doorbell registers and PCIe BARs), and connect ULL-Flash to DRAM PHY (instead of PCIe). 
Note that writes to ULL-Flash are already reduced without employing the SSD-internal DRAM  as the incoming data are buffered/cached by NVDIMM. Similarly, the address mapping table is also buffered in the NVDIMM. Accessing the mapping information only consumes a \emph{tCL} and a few \emph{tBURST} periods (less than 20ns), which is ignorable compared to the long ULL-Flash access latency.
As shown in Figure \ref{fig:over_TX}, this aggressive integration of NVDIMM and ULL-Flash, which we call advanced HAMS, allows the NVMe controller to directly access the DRAM modules over the DRAM interface. 
Specifically, to be compatible with the synchronous DDR4 interface, the NVMe controller avoids unpredictable delay of the underlying Z-NAND accesses by employing a set of registers to buffer the command, address, and data. For communications, the address manager employs an SSD command generation logic that writes a set of registers capturing the source and destination addresses and I/O command, based on the I/O request that HAMS needs to initiate. 
The NVMe controller fetches (or pulls) the target data from the source address of NVDIMM (written to the address register via the DRAM interface) 
and then forwards it to flash firmware so that it can be programmed into flash media. 

While allocating multiple DDR4 channels to connect each pair from HAMS controller, NVDIMM and ULL-Flash can parallelize the MMU operations and ULL-Flash read/write accesses, this design also makes DDR4 channels under-utilized. To avoid wasting the channel resources, we propose to connect ULL-Flash and one/multiple NVDIMMs to HAMS controller via the same DDR4 bus. However, one of the key design issues is that NVDIMM can be accessed by both the HAMS controller and NVMe controller in our design. 
To avoid simultaneous accesses from these two, this aggressive integration also introduces a \textit{lock register}, which indicates that the NVMe controller is in the process of accessing DDR4 and NVDIMM for data transfers.

\begin{figure}
\centering
\includegraphics[width=0.9\linewidth]{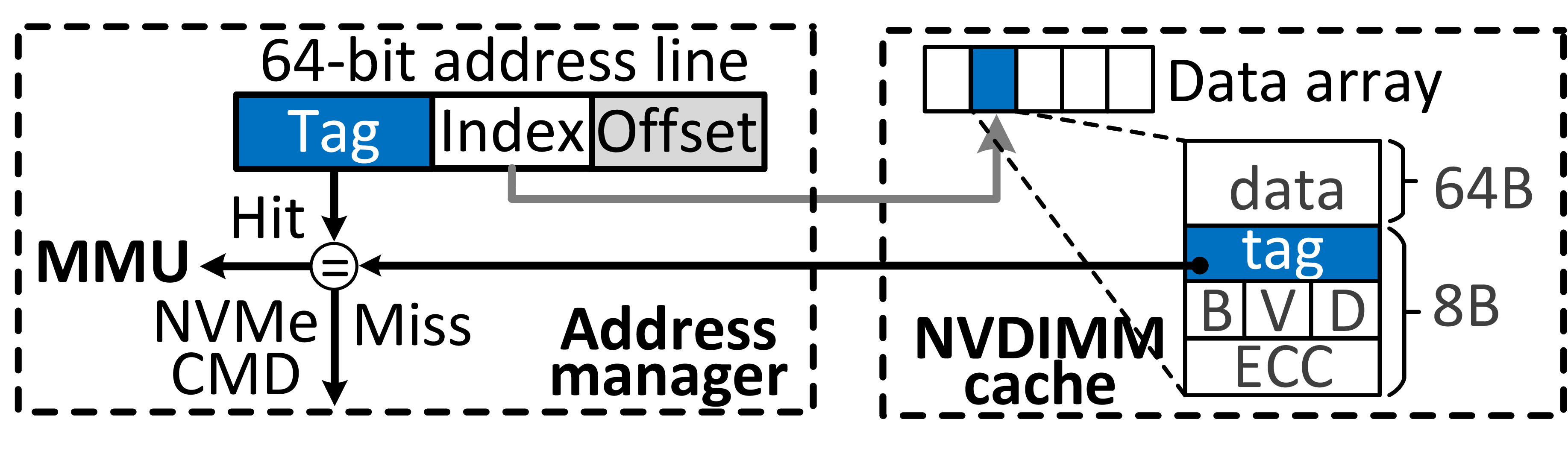}
\caption{\label{fig:impl_init2}MoS tag-array design in NVDIMM cache.\vspace{-5pt}}
\vspace{-15pt}
\end{figure}

\section{Implementation Details}
\label{sec:Implementation}

\subsection{NVDIMM Cache and Bus Integration}
\label{sec:addrmanage}
\noindent \textbf{HAMS address management.} 
An SRAM-based MoS tag-array can expose a significant circuit area cost to the HAMS controller and raise the concern of metadata persistency when a power failure occurs. 
Instead, we configure MoS' NVDIMM cache as direct mapped and integrate its tag information along with ECC bits in each NVDIMM cache line, which is similar to the MCDRAM configuration of Intel Knights Landing processor \cite{sodani2016knights}. Figure \ref{fig:impl_init2} shows details of our MoS tag-array in the NVDIMM cache. Each entry of the MoS tag-array contains all metadata of the cache, such as the tag, busy bit (B), valid bit (V) and dirty bit (D). When there is an incoming memory request, its address is decomposed into the tag, index, and offset fields. 
HAMS address manager then retrieves the tag-array entry and the data block from the NVDIMM cache by using the decomposed index. A comparator pulls the stored tag from the retrieved tag-array entry and compares it with the tag of the corresponding memory request. 
If the two tags match, the fetched data can be directly served from HAMS controller. 
On the other hand, if the two tags mismatch, HAMS composes two NVMe commands, one for a read that fills data from ULL-Flash to the NVDIMM cache entry, and another for a write that evicts the data from NVDIMM to ULL-Flash. Once the target data are available in the NVDIMM cache, HAMS places it on the system bus, and notifies the completion to CPU by setting the MMU's command and address buses.

\ignore{
I think THAMS controller follows the normal process to interact with cache. 
1. If there is a read cache miss, cache controller puts the request information in MSHR and sends the requests to THAMS controller via command bus and address bus.
2. THAMS controller parses the incoming requests and fetches data from DRAM chip. 
3. THAMS controller puts the data in data bus and set command bus and address bus appropriately. 
3. Once cache receives the response, it clears the record in MSHR and transfer the data to register files. Then it informs CPU to continue execution.
}

\begin{figure}
\centering
\includegraphics[width=1\linewidth]{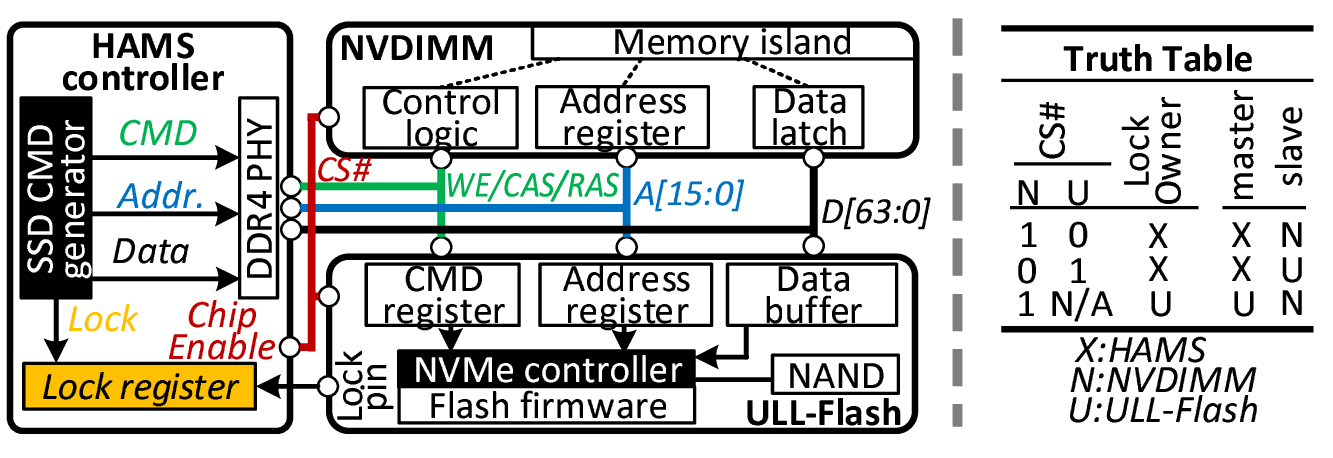}
\vspace{-15pt}
\caption{\label{fig:impl_TX}The details of the register-based interface.\vspace{-5pt}}
\vspace{-15pt}
\end{figure}

\noindent \textbf{Register-based interface.} Figure \ref{fig:impl_TX} illustrates how our advanced HAMS controller communicates with the underlying NVMe controller through DDR4.
In our design, to send an I/O request to NVMe controller, the HAMS controller firstly deselects the NVDIMM by toggling its \textit{CS\#} strobe to high voltage. In the next clock cycle, the HAMS controller configures the write command in the DDR channel by toggling the \textit{WE\#}, \textit{CAS\#} and \textit{RAS\#} strobe to low, low and high voltage, respectively. Following the write command, the I/O request, which is packeted as a 64B NVMe command, is transferred to the data buffer registers of ULL-Flash via the \textit{D[63:0]} strobes in 8-cycle data burst. The NVMe controller then extracts and parses the request information (i.e., request type, source/destination addresses and data length) from the data buffer registers, similar to most NVMe SSDs. Note that, unlike their original purpose, the address strobes \textit{A[15:0]} deliver no information during the communication between HAMS controller and NVDIMM controller.

After a given number of cycles for processing the NVMe write command or fetching data from the flash media based on NVMe read command, HAMS sets the lock register to 1, which indicates that the NVMe controller can take over the control as a bus master. 
If the lock register is configured, the NVMe controller initializes the DMA procedure between ULL-Flash and NVDIMM based on the timing sequence of the DDR4 interface. 
After transferring data, the NVMe controller releases the lock register by resetting it to 0.
HAMS cache logic uses the lock register for NVDIMM accesses, which helps us avoid a case where both NVMe controller and memory controller use the bus at the same time.
 

\subsection{NVMe and Hazard Management}
The I/O requests for each NVMe queue entry can be simply composed by filling the information fields of the NVMe command structure. 
HAMS writes the opcode field for a given request (read/write), and fills the NVDIMM address, SSD address, and page size (4KB) into the corresponding PRP, LBA, and length fields, respectively. 
The generated NVMe command is enqueued in the SQ by the HAMS NVMe engine. 
This engine writes the doorbell to inform the ULL-Flash of a request. 
Whenever the interrupt is delivered from the ULL-Flash controller of HAMS, 
the NVMe engine synchronizes the corresponding CQ and clears the target entries of CQ and SQ. 

\begin{figure}
\centering
\includegraphics[width=0.95\linewidth]{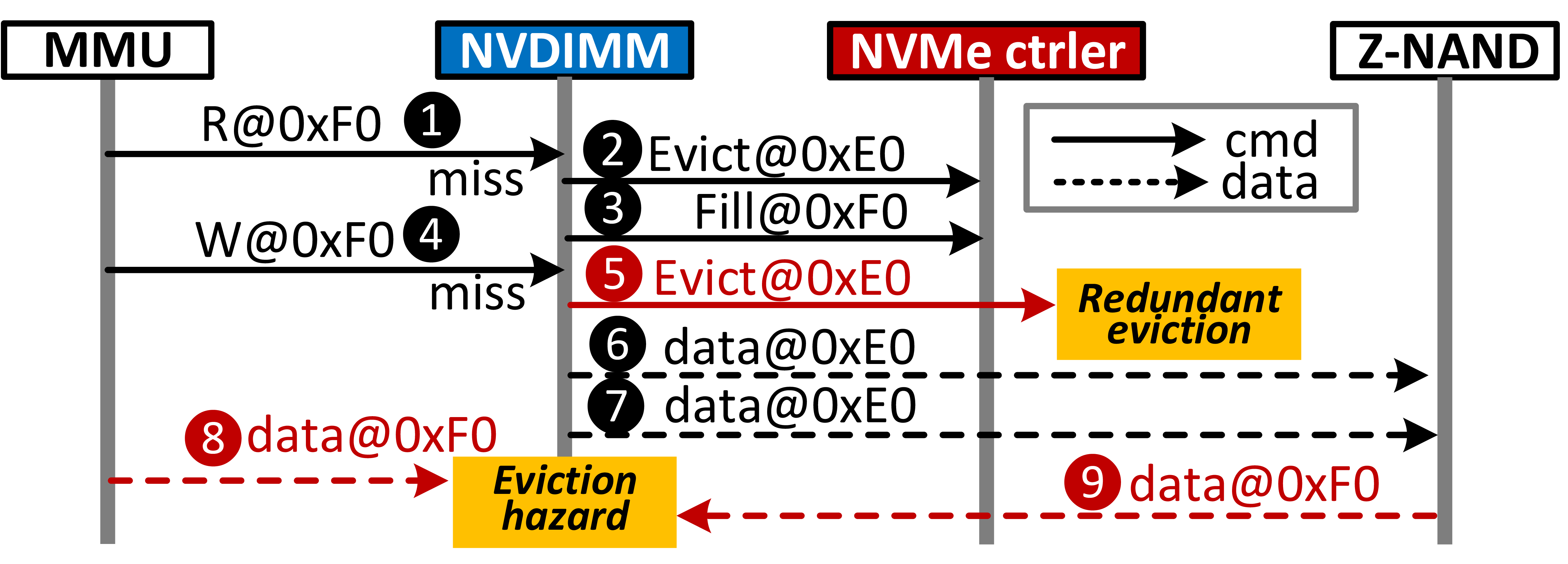}
\caption{\label{fig:over_LX_chal}Challenges with the baseline HAMS.\vspace{-5pt}}
\vspace{-10pt}
\end{figure}

There are two issues associated with this NVMe management, as NVDIMM is used both as a cache and as a PRP target: (1) \textit{eviction hazard} and (2) \textit{redundant eviction}. 
The eviction hazard occurs when the NVMe controller and HAMS cache logic access the same NVDIMM location, 
whereas the redundant eviction arises when the cache logic generates an eviction command, which is already being issued. 
Consider the example illustrated in Figure \ref{fig:over_LX_chal}. 
The MMU requests a read at \texttt{0xF0} of the MoS address space, the index and tag of which are \texttt{0x0} and \texttt{0xF}, respectively. 
Since a cache miss occurs, the HAMS cache logic evicts the exiting page (\texttt{0xE0}) to ULL-Flash, and requests a data read at \texttt{0xF0}. 
In the meantime, the MMU accesses \texttt{0xF0} to update the data. 
This makes the cache logic evict the same data again, because the evicted request is still serviced by ULL-Flash (i.e., redundant eviction). 
Now, the HAMS NVMe engine contains three NVMe commands (\texttt{CMD1/2/3}). 
These commands are processed by the NVMe controller in a FIFO order based on the NVMe specification. 
However, I/O completions within ULL-Flash can be out-of-order, due to SSD-internal tasks. 
More importantly, the NVMe controller transfers the data to NVDIMM based on the order of completion, which can cause an eviction hazard.

\begin{figure}
\centering
\includegraphics[width=0.95\linewidth]{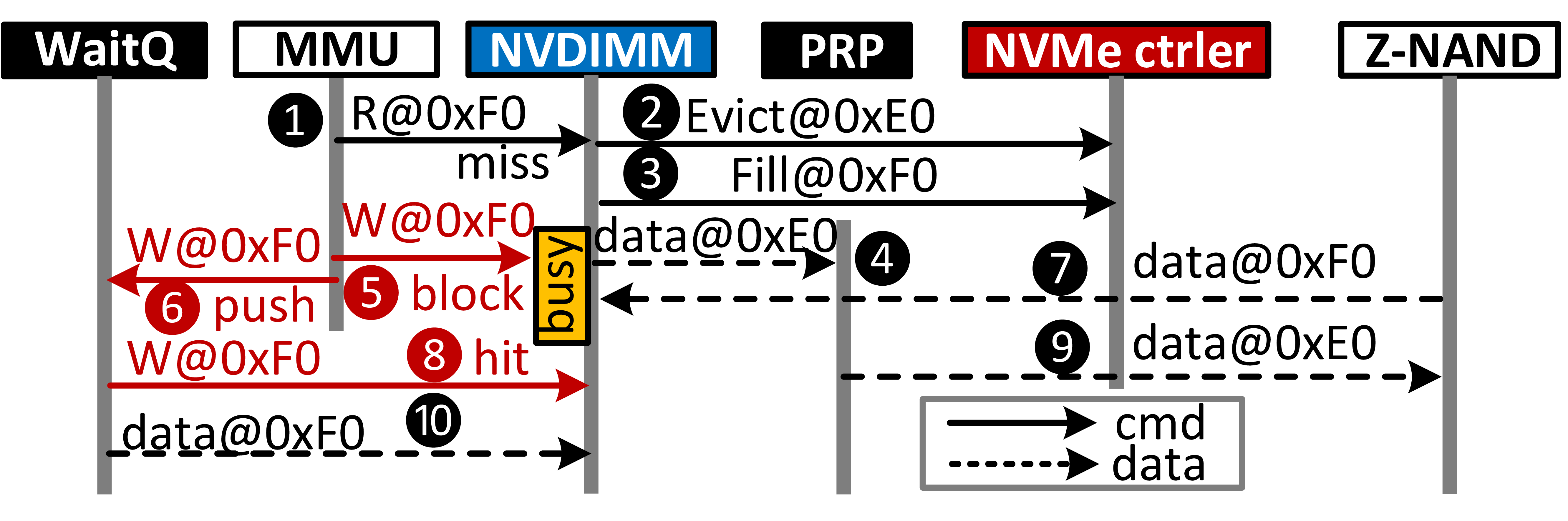}
\vspace{-5pt}
\caption{\label{fig:impl_access}Hazard avoidance methods.\vspace{-10pt}}
\vspace{-5pt}
\end{figure}

To prevent these hazards and redundant evictions, HAMS employs two techniques. 
When the NVMe engine issues commands, HAMS isolates the target contents from the corresponding NVDIMM cache entry by cloning the corresponding page into the PRP pool allocated in the pinned memory (Figure \ref{fig:impl_init1}). 
It then updates the PRP value with the location of the cloned page so that the underlying NVMe controller does not make the data inconsistent during DMA. 
Further, we add a wait queue to the pinned memory, and make HAMS always refer to a busy bit (cf. Section \ref{subsec:manage}) of the MoS tag-array, 
whenever a cache miss occurs. 
The HAMS cache logic sets the bit to 1 and then resets it to 0, when the NVMe engine completes the request. 
Figure \ref{fig:impl_access} shows an example that illustrates how the eviction hazard and redundant eviction issues are handled. 
When a cache miss occurs (read at \texttt{0x0E}), the cache logic toggles the busy bit of the target tag-array's entry and copies the target page to a PRP pool entry. 
During this process, HAMS replaces the reference to PRP with the PRP pool entry and submits it to the NVMe engine. 
Upon the next cache miss (write to \texttt{0xF0}), the cache logic realizes that the entry is in an eviction process, and puts the request into the wait queue. 
After the I/O services of the NVMe commands are completed, the busy bit is cleared, and the request that sits in the wait queue is issued again. 
In this way, the eviction hazard and redundant eviction issue with the wait queue and busy bit can be avoided.

\subsection{Persistency Control Upon Power Failure}
A target system can benefit from a large memory space, if it utilizes HAMS as a working memory expansion, which expands the address space by combining NVDIMM and ULL-Flash. 
However, it needs a guarantee for \textit{data persistency}, as MoS address space is considered as a persistent memory expansion.
Thus, HAMS requires to flush the NVMe request whenever its cache logic needs to update data in ULL-Flash. 
To address this shortcoming, we add a \textit{journal tag} to each SQ's NVMe command entry by utilizing the reserved area in the NVMe command format. 
This journal tag keeps information that indicates whether the corresponding request is completed by ULL-Flash. 
Whenever the NVMe engine sends a request to ULL-Flash, it sets the tag to 1. 
Once the interrupt arrives, HAMS clears the tag associated with the I/O completion. 

\begin{figure}
\centering
\includegraphics[width=0.9\linewidth]{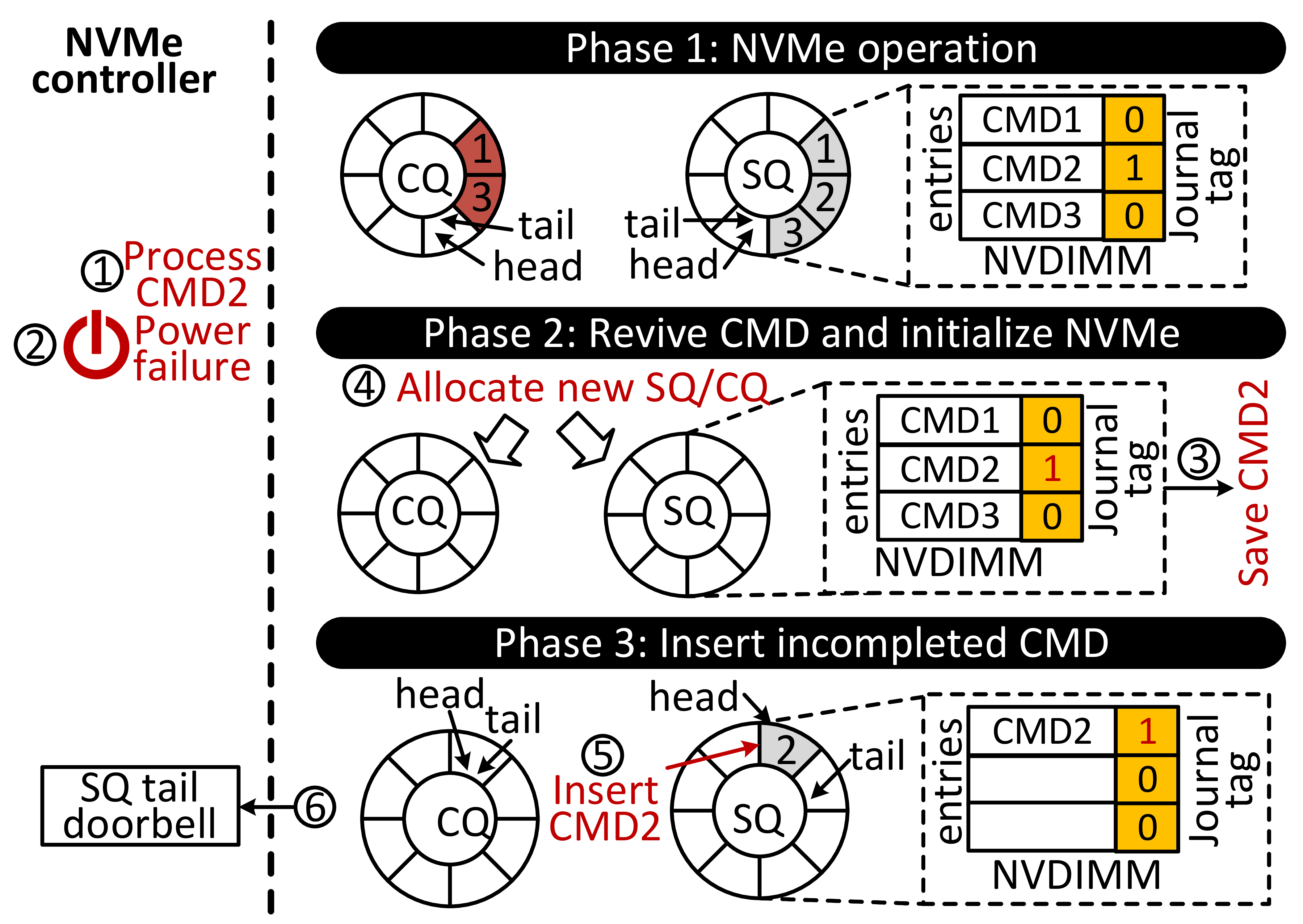}
\caption{\label{fig:impl_power}Power failure recovery procedure.\vspace{-5pt}}
\vspace{-15pt}
\end{figure}

Figure \ref{fig:impl_power} gives an example that illustrates how HAMS utilizes the journal tag information. 
In the first phase of this example, HAMS issues all the commands in the SQ to ULL-Flash, and \texttt{CMD1}, \texttt{CMD3} and \texttt{CMD4} are processed 
as the tail and head pointers refer to the same location in the SQ/CQ, which clears the corresponding journal tags to 0. 
Upon a power failure at the end of the first phase, ULL-Flash and HAMS cannot finish \texttt{CMD2}. 
Since the pinned memory space of NVDIMM holds the data of the SQ region in our design, HAMS first checks the SQ region on power-up to determine if there is any command whose journal tag is 1. 
If there is one, HAMS pulls the command and creates a pair of SQ and CQ for the I/O service in second phase. 
HAMS then restores it to the SQ, increases the SQ's tail pointer, and rings the doorbell register, so that the outstanding request issued at the moment of a power failure can be served appropriately.

\begin{table*}[]
\centering
\begin{minipage}{.28\linewidth}
\resizebox{\textwidth}{!}{
\begin{tabular}{|l|c|}
\hline
\textbf{OS} & Linux 4.9, Ubuntu 14.10 \\ \hline
\textbf{CPU} & quad-core, ARM v8, 2GHz \\ \hline
\textbf{Cache} & 64KB L1I/64KB L1D/2MB L2 \\ \hline
\textbf{memory} & NVDIMM, DDR4, 8GB, \newedit{128KB page} \\ \hline
\textbf{storage} & ULL-Flash, 512MB buffer, 800GB \\ \hline
\textbf{flash} & 3us read, 100us write \\ \hline
\end{tabular}}
\caption{Gem5 specification. \label{table:gem5}}
\end{minipage}%
\begin{minipage}{.7\linewidth}
\resizebox{\textwidth}{!}{
\begin{tabular}{|l|c|c|c|c|c|c|c|c|c|c|c|c|}
\hline
\textbf{Benchmark} & \multicolumn{4}{c|}{\textbf{Microbenchmark}} & \multicolumn{5}{c|}{\textbf{SQLite benchmark}} & \multicolumn{3}{c|}{\textbf{Rodinia}}\\ \hline
\textbf{Workloads} & \multicolumn{1}{l|}{seqRd} & \multicolumn{1}{l|}{rndRd} & \multicolumn{1}{l|}{seqWr} & \multicolumn{1}{l|}{rndWr} & \multicolumn{1}{l|}{seqSel} & \multicolumn{1}{l|}{rndSel} & \multicolumn{1}{l|}{seqIns} & \multicolumn{1}{l|}{rndIns} & \multicolumn{1}{l|}{Update}  & \multicolumn{1}{l|}{BFS} & \multicolumn{1}{l|}{KMN} & \multicolumn{1}{l|}{NN}\\ \hline
\textbf{\# of inst.} & 67G & 69G & 67G & 69G & 213G & 213G & 40G & 44G & 244G & 192G & 38G & 145G \\ \hline
\textbf{load inst. ratio} & 0.28 & 0.27 & 0.28 & 0.27 & 0.26 & 0.26 & 0.25 & 0.25 & 0.26 & 0.21 & 0.27 & 0.16\\ \hline
\textbf{store inst. ratio} & 0.43 & 0.37 & 0.43 & 0.37 & 0.20 & 0.20 & 0.21 & 0.21 & 0.20 & 0.04 & 0.03 & 0.05\\ \hline
\textbf{Data sets} & 16GB & 16GB & 16GB & 16GB & 11GB & 11GB & 11GB & 11GB & 11GB & 9GB & 5GB & 7GB\\ \hline
\end{tabular}}
\caption{Workload characteristics. \label{tab:workload-charac}}
\end{minipage}
\vspace{-10pt}
\end{table*}


\ignore{
\noindent \textbf{Synchronization of pinned NVDIMM region.}
2.   NVMe engine: wait unitl CSTS.RDY = 0
3.   NVMe engine: Write AQA to set entry size of Admin submission completion queue
4.   NVMe engine: Write ASQ to indicate memory base address of SQ
5.   NVMe engine: write ACQ to indicate memory base address of CQ
6.   NVMe engine: enable controller by writing 1 to CC.EN.
7. NVMe engine: wait unitl CSTS.RDY = 1.
}

\ignore{
\section{Modeling NVMe and ULL-Flash}

Existing SSD simulators unfortunately lack hardware and/or software architecture models for SSD external and internal datapath. Consequently, they are far from capturing the critical features of ULL SSD devices. More importantly, traditional SSD simulators \cite{} capture only storage-related metrics such as bandwidth and latency by replaying block-level I/O traces; this ignores system-level interaction between the host-side CPU and storage subsystems. 

To address these challenges, we propose a high-fidelity simulator that models all of the detailed characteristics of hardware
and software of NVMe SSD internals. We further modify the I/O bridge of gem5 by incorporating our NVMe SSD simulation framework so that it can accommodate complete storage stack of a real Linux system. 


\noindent \textbf{Software model.}
One of the main challenges of simulating an SSD is supporting diverse flash firmware versions, which greatly influences the target storage performance. We model a flexible flash translation layer (FTL) whose address translation mechanism can simply be reconfigured based on different associativity granularities defined by system architects. We also decouple I/O scheduling and page allocation mechanisms from the FTL so that new scheduling proposals that are aware of SSD-internal parallelism can be embedded without changing the FTL. Although we do not cover all types of potential FTLs, the implemented reconfigurable mapping algorithm can capture/support diverse operational characteristics of a block level mapping FTL, a fully-associative FTL, and various hybrid mapping schemes that employ different levels of block and page mapping tables in their address translations. In addition, our simplified but reconfigurable layered firmware also offers diverse research opportunities where system and computer architects can simply modify some performance-critical components such as garbage collection and wear-leveling algorithms with different mapping mechanisms.

\noindent \textbf{Hardware.} The performance characteristics of the underlying hardware vary based on i) the intrinsics of latency of individual flash characteristics and ii) their different levels of parallelism. A cycle-level simulation for each component can accurately evaluate all SSD internals. However, full-system simulations with an SSD at the cycle level require an unreasonably long runtime and excessive resources. In this work, we abstracted both flash-level and subsystem-level hardware characteristics. We implemented an FPGA-based memory controller built on Xilinx Spartan-6 and then used this to characterized different memory technologies. Based on the extracted characteristics, we first design a die-level latency model by simplifying the flash transactions. Specifically, we examined all flash transactions specified by the open NAND flash interface (ONFi 3.x \cite{}) and classified various timing components of the corresponding protocol into a few transaction activities. With this simplified latency model, the proposed SimpleSSD simulates varying numbers of flash chips over many interconnection buses by modeling the executions across different hardware resources and resource contentions. Even though this simplified model cannot account for all of the characteristics from the flash at a cycle level, it can capture the close interactions among the designs of the firmware, controller, and architecture by being aware of flash latency intrinsics and internal parallelism. 

}

\section{Evaluation}
\label{sec:evaluation}
\subsection{Experiment Setup}
\label{sec:experiment}
\noindent \textbf{Simulation model.}
To explore the full design space of the HAMS enabled systems from both the software and hardware perspectives, we first replace the existing main memory implementation in a full system simulator (gem5 \cite{binkert2011gem5}) with the latency model of an 8GB DRAM-based NVDIMM \cite{nvdimm8G}. We then model an ULL-Flash archive and integrate it into gem5 by revising the memory controller and I/O bridge model\footnote{All source codes of our full-system simulation that integrates high-fidelity SSD storage models will be made available for download in public domain.}.
The storage-side components of the proposed simulator are configured as ULL-Flash instances by leveraging an existing SSD simulator, Amber \cite{gouk2018amber}, which is highly reconfigurable (being aware of the details of flash internals, SSD internals and parallelism-related design parameters) and detailed (implementing a full firmware stack and an actual NVMe interface). Our simulator has been verified with an actual 800GB ULL-Flash prototype \cite{samsung2017znand}. 
Note that this proposed simulation framework enables the execution of data-intensive applications on a real Linux, while allowing us to investigate the full design space on the datapath from top to bottom. The details of our simulation environment are given in Table \ref{table:gem5}.

\noindent \textbf{Experiment precondition and energy profiling method.} To guarantee the consistency of our experimental results, we completely wrote all data-blocks into the flash-media, and flushed/cleaned up the internal-DRAM in a \textit{warm-up phase} before performing our evaluations. The energy estimation of each component in the full-system platform is performed based on its power model; more specifically, the power models of ULL flash and NVDIMM are derived based on NAND flash datasheets and MICRON SDRAM power calculator \cite{sdrampower}, which will be available for download (along with our simulator), while the energy consumptions of core and cache are measured by leveraging McPAT \cite{li2009mcpat}.

\noindent \textbf{Benchmarks.} We evaluate 12 data-intensive workloads from MMF microbenchmark \cite{matt2014mmap}, Rodinia \cite{che2009rodinia}, and SQLite \cite{google2014} benchmarks. 
While MMF microbenchmark is memory-intensive, Rodinia benchmark requires high computation.
In addition, MMF microbenchmark accesses the persistent memory system in a coarse-granular fashion (i.e., by pages). In contrast, the other workloads generate fine-granular memory accesses ranging from 8B to 100B. In our experiments, the datasets to be tested initially reside in either ULL-Flash or HAMS. To access data, these workloads are structured to support memory-mapped file I/O via the POSIX-compliant system call \texttt{mmap}. Table \ref{tab:workload-charac} tabulates the important characteristics of our benchmarks such as the total number of instructions, fraction of load/store instructions, and dataset sizes.

\noindent \textbf{Simulation platforms.} We configured a traditional computer system, called \texttt{mmap}, as our baseline for evaluation. \texttt{mmap} employs an ULL-Flash and a DDR4 DRAM as its storage and memory media, respectively. Table \ref{table:gem5} shows important parameters of our system configurations. By default, the baseline accesses data directly from the persistent storage by using the MMF module. We also built five computing platforms employing the existing memory expansion techniques \cite{lee2020nvdimm, abulila2019flatflash, yang2020empirical} and four different systems that implement our \texttt{HAMS} model. Specifically, (1) \texttt{optane-P} \cite{optane-perf} employs 512GB Optane DC PMM as main memory. To guarantee data persistency, Optane DC PMM operates in App Direct mode that serves all memory requests without DRAM cache. (2) \texttt{optane-M} \cite{optane-perf} employs 8GB DRAM as the cache of Optane DC PMM, which can improve the performance but sacrifice the data persistency. (3) \texttt{flatflash-P} \cite{abulila2019flatflash} allows the applications to directly access a cache line from ULL-Flash via MMIO \cite{bae20182b} thereby guaranteeing the data persistency. (4) Compared to \texttt{flatflash-P}, \texttt{flatflash-M} \cite{abulila2019flatflash} selectively buffers hot pages in 8GB host-side memory for fast accesses. (5) \texttt{nvdimm-C} \cite{lee2020nvdimm} connects ULL-Flash to DRAM PHY thereby sharing the memory channel with DRAM. \texttt{nvdimm-C} uses DRAM as a cache of ULL-Flash. However, data migration between DRAM and ULL-Flash is only allowed during DRAM refresh periods. (6) A loosely-coupled HAMS system, which connects to 8GB NVDIMM and 800GB ULL-Flash via a memory channel and PCIe links, respectively, is referred to as \texttt{hams-L}. \texttt{hams-LP} is the loosely-coupled HAMS system, which works in a ``persist mode'' to persistently store data. \texttt{hams-LP} tags FUA per I/O request and enforces at most a single I/O request on-the-fly. (7) \texttt{hams-LE} is also the loosely-coupled \texttt{HAMS} system, but it operates in an ``extend mode''. In particular, \texttt{hams-LE} leverages the NVMe protocol to enable parallel accesses to ULL-Flash. To guarantee the data persistency, it also employs our proposed persistency control to manage power failure. (8) An advanced HAMS system with aggressive integration is referred to as \texttt{hams-T}. \texttt{hams-TP} employs such HAMS system, which works on persist mode. Lastly, (9) \texttt{hams-TE} employs \texttt{hams-T}, but the extend mode. 
\newedit{Lastly, we configure an \texttt{oracle} platform that employs a 512GB NVDIMM to serve the evaluated workloads.}

\begin{figure}
\centering
\subfloat[MMF and Rodinia benchmarks.]{\label{fig:app1}\rotatebox{0}{\includegraphics[width=1\linewidth]{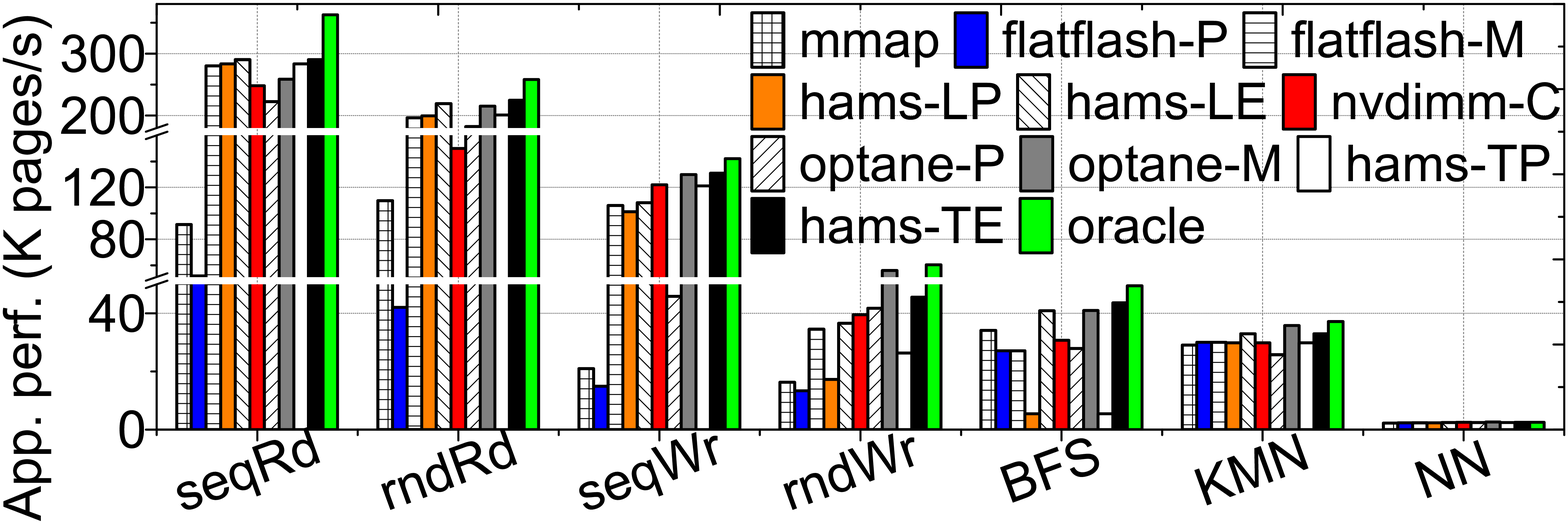}}}

\vspace{-10pt}
\subfloat[SQLite benchmark.]{\label{fig:app2}\rotatebox{0}{\includegraphics[width=1\linewidth]{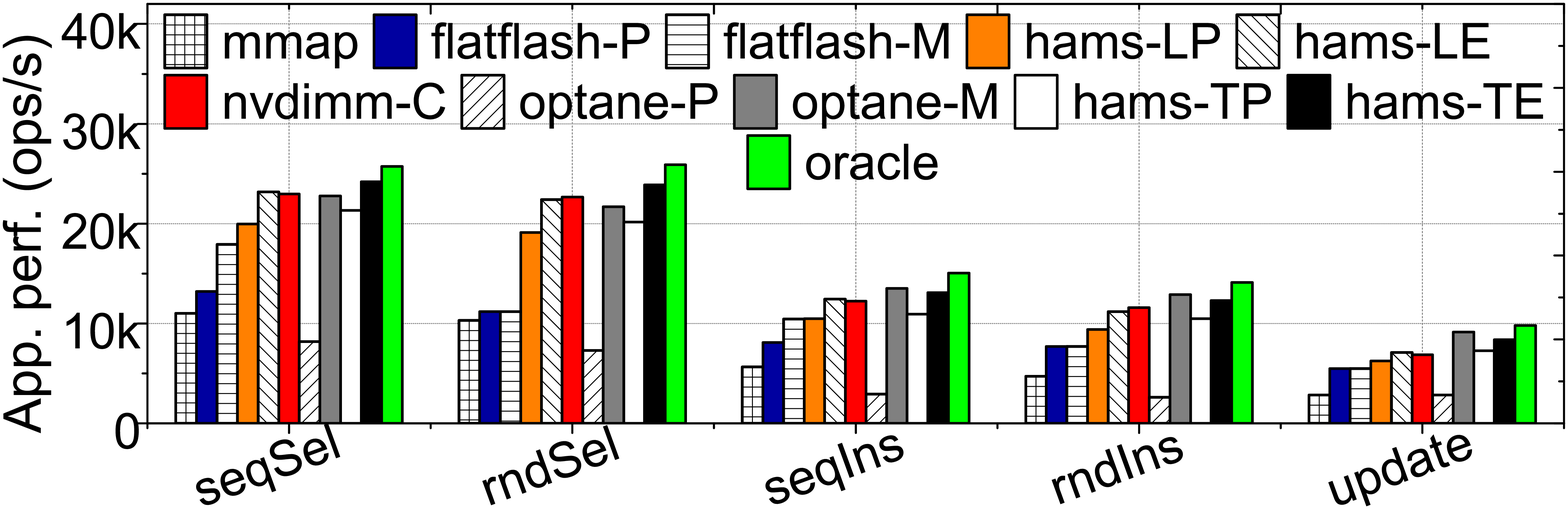}}}
\caption{\label{fig:bck_overall}Application performance.\vspace{-5pt}}
\vspace{-15pt}
\end{figure}


\subsection{System Performance Analysis}
\label{sec:performance_analysis}

\noindent \textbf{Application-level performance.}
Figures \ref{fig:bck_overall} plots the performance for microbenchmark, Rodinia, and SQLite benchmark. \texttt{mmap} achieves 43K pages/s for the microbenchmark and graph workloads, and 6905 SQL ops/s for the SQLite workloads, on average, which are, respectively, 2.54x and 1.37x worse than \texttt{hams-TE}. This is because, the I/O requests in \texttt{mmap} go through a complex system software stack before finally reaching the storage, which introduces a substantial overhead.
Although \texttt{flatflash-P} allows CPU to directly access data from the storage, it consumes 4.8us for 64B data access, which is over 40 times longer than DRAM access latency \cite{abulila2019flatflash}. Thus, \texttt{flatflash-P} degrades performance by 75\% in the workloads of MMF microbenchmark, compared to \texttt{mmap}. Since \texttt{flatflash-M} can hide the long storage access latency by buffering hot pages in host-side memory, \texttt{flatflash-M} outperforms \texttt{flatflash-P} by 136\%, on average, in all evaluated workloads. However, \texttt{flatflash-M} accesses the storage via MMIO rather than the NVMe protocol, which loses the opportunity of utilizing the plenty of queue and flash parallelisms. In contrast, \texttt{hams-LE} implements NVMe protocol in our HAMS controller to enable parallel accesses to the underlying ULL-Flash. In addition, \texttt{hams-LE} mitigates the storage access overhead from OS by offloading the task of page access to hardware. Therefore, \texttt{hams-LE} improves the performance by 26\% in all the workloads, on average, compared to \texttt{flatflash-M}.
\texttt{nvdimm-C} further improves the efficiency of the storage access by directly connecting ULL-Flash to the host-side DRAM via the same memory channel. However, to prevent the memory controller and SSD controller from competing for the memory channel, \texttt{nvdimm-C} constrains the data migration between DRAM and ULL-Flash in the period of DRAM refresh operations. Although fetching a single page from ULL-Flash costs 3us, moving data from ULL-Flash to DRAM consumes upto 48us \cite{lee2020nvdimm}. 
Considering such long latency, it is difficult to execute latency-critical applications in \texttt{nvdimm-C}. In contrast, HAMS can fit to a wider range of applications, especially the ones with large memory footprint.
\texttt{optane-P} outperforms \texttt{mmap} by 121\% in microbenchmark, as all data initially reside in the persistent memory, which eliminates the overhead of moving data between memory and storage. However, the performance of \texttt{optane} is unfortunately not promising in the workloads with fine-granular memory accesses (i.e., Rodinia and SQLite benchmark). This is because the memory request size is much smaller than the internal block size (256B) of Optane DC PMM, which wastes the memory bandwidth. \texttt{optane-P} resolves the mismatch between the memory request size and the Optane internal block size by employing DRAM as the cache of Optane DC PMM. Such design improves the performance by 142\%, compared to \texttt{optane-P}. On the other hand, \texttt{hams-T} serves the memory requests from NVDIMM, whose bandwidth is not constrained by the internal block size. 
In addition, \texttt{hams-T} enables direct access between NVDIMM and ULL-Flash, which eliminates the redundant data copies. Thus, \texttt{hams-TE} improves the application's performance by as high as 12\%, compared to \texttt{optane-M}.
\newedit{Lastly, as the data migration latency cannot be fully overlapped with computation time in data-intensive workloads (e.g., \emph{seqRd} and \emph{seqWr}), \texttt{hams-TE} performs worse than \texttt{Oracle} by 14\%.}

\begin{figure}
\centering
\includegraphics[width=1\linewidth]{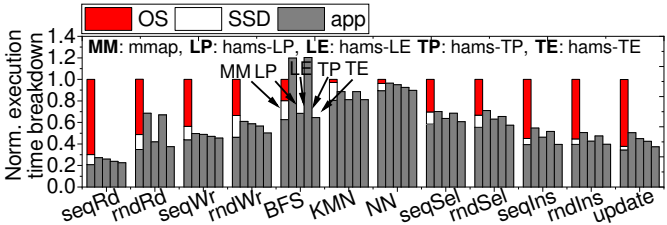}
\vspace{-10pt}
\caption{\label{fig:exe-break}System-level execution time breakdown.}
\vspace{-25pt}
\end{figure}

\noindent \textbf{Execution time breakdown.}
We now break down the execution time of our workloads from the view point of system software, to analyze the critical factors that can impact the overall performance. Figure \ref{fig:exe-break} shows the execution time breakdown. As shown in the figure, in \texttt{mmap}, a large fraction of the execution time is consumed by the ``OS'' and ``SSD'' accesses. The overheads brought by the ``OS'' and ``SSD'' accesses cannot be hidden by application execution, as the application is always stalled until the OS fetches data from storage and prepares it in the main memory. Since we are using an ultra-low latency SSD (ULL-Flash), the overheads brought by the storage accesses are not the main factor that degrades the overall performance. Instead, as current Linux kernel is \textit{not} optimized for ULL-Flash, it becomes a performance bottleneck in the baseline platform. On the other hand, the overheads brought by ``OS'' and ``SSD'' can be ignored in \texttt{HAMS}, as \texttt{HAMS} hybrids NVDIMM and ULL-Flash in the main memory, and directly accesses ULL-Flash as memory without any OS intervention. Note that the storage-access times are excluded from ``app'' and separately presented with the labels of ``OS'' and ``SSD'', whereas the storage-access times of \texttt{HAMS} are included in ``app'' (as they are classified as the latencies of LD/ST instructions). The ``app'' time of \texttt{hams-TE} is as short as that of \texttt{mmap}, indicating that \texttt{hams-TE} can \textit{fully hide} the OS and SSD access overhead. 

\subsection{Detailed Analysis}

\begin{figure}
\centering
\includegraphics[width=1\linewidth]{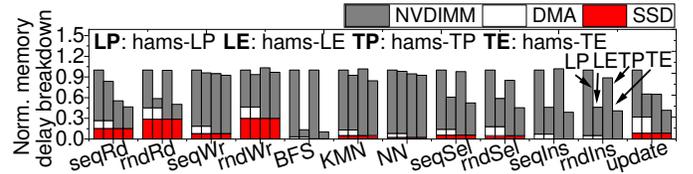}
\vspace{-10pt}
\caption{\label{fig:mem-break}Memory access delay breakdown.}
\vspace{-15pt}
\end{figure}

\noindent \textbf{Memory latency analysis.} We collect the statistics from the memory-side and present the hardware performance in terms of memory stalls in Figure \ref{fig:mem-break}. 
As our HAMS employs a large NVDIMM as cache (i.e., 8GB in Table \ref{table:gem5}) to accommodate most memory requests, the cache hit rate of NVDIMM reaches 94\%, on average, in all the tested workloads. Thus, NVDIMM accesses account for 79\% of the total memory delay in \texttt{hams-LP}.
\texttt{hams-T} (including \texttt{hams-TP} and \texttt{hams-TE}) reduces the total memory stalls by 16\%, compared to \texttt{hams-L}. This is because \texttt{hams-T} leverages the DDR4 interface to directly transfer data between NVDIMM and SSD while \texttt{hams-L} uses different interfaces (DDR and NVMe/PCIe) for NVDIMM and ULL-Flash, and always requires extra time to transform data format. On the other hand, the persist mode generates 34\% more memory delay than the extend mode, on average. This is because, the persist mode only allows one memory access at a time, which means serializing the executions of instructions that experience cache misses. For \texttt{hams-L}, NVMe-DMA contributes to 18\% of the memory delay in data-intensive workloads such as \emph{rndRd}, \emph{rndWr}, \emph{seqRd}, \emph{seqWr} and \emph{update}. This is because, the PCIe link used by NVMe SSD is mainly designed for peripheral devices and provides much lower bandwidth compared to the DDR4 interface. Thus, transferring data via PCIe costs much longer time than the DDR4 interface. On the other hand, for other workloads that do not intensively access the storage, \texttt{hams-L} and \texttt{hams-T} have similar memory delays. 

\begin{figure}
	\centering
	\includegraphics[width=1\linewidth]{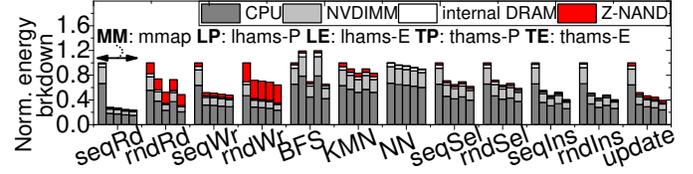}
	\vspace{-10pt}
	\caption{\label{fig:energy_break}Energy breakdown normalized to \texttt{mmap}.}
	\vspace{-20pt}
\end{figure}


\noindent \textbf{Energy analysis.}
Figure \ref{fig:energy_break} plots the energy consumption of the \textit{whole system} including CPU, system memory (DRAM), ULL-Flash internal DRAM, and Z-NAND chips. As shown in the figure, \texttt{hams-LP}, \texttt{hams-LE}, \texttt{hams-TP}, and \texttt{hams-TE} reduce the system-level energy by 31\%, 41\%, 34\%, and 45\%, respectively, compared to \texttt{mmap}. Specifically, the combined energy of CPU and system memory in \texttt{mmap} is 89\% higher than that of \texttt{hams}, as \texttt{mmap} spends more time for I/O responses, which also costs more CPU and memory idle energy. On the other hand, the persist mode and extend mode do not impact the energy consumption of NVDIMM. This is because, the persist mode only constrains the number of memory requests on the fly, but it does not impact the total number of NVDIMM accesses. In contrast, \texttt{hams-L} consumes 8\% more NVDIMM energy than \texttt{hams-L}. This is because, \texttt{hams-T} directly transfers data between NAND flash and NV-DIMM without any redundant data copies, whereas \texttt{hams-L} employs NVDIMM and SSD internal DRAM to buffer data and this introduces redundant data copies. 
\texttt{HAMS} also reduces the energy of accessing SSD by 11\%, compared to \texttt{mmap}, on average. This is because, \texttt{mmap} needs to periodically flush data from the main memory to SSD for persistency. 

\noindent \textbf{Overhead analysis.}
Compared to the existing memory controller, \texttt{HAMS} requires NVMe queue engine, SSD command generator and lock register. In our design, 
the core logic of the NVMe queue engine and SSD command generator employ thousands of gates, which are negligible compared to the billion transistors in modern CPUs.
While HAMS enables ULL-Flash to share the DDR4 channels with the NVDIMMs to avoid extra usage of channel resources, ULL-Flash can occupy one DIMM slot. However, considering the fact that LRDIMM supports up to 24 DIMM slots, HAMS only reduces the maximal memory capacity by 4\%.

\begin{figure}
\centering
\vspace{-10pt}
\subfloat[Various page sizes.]{\label{fig:pagesize}\rotatebox{0}{\includegraphics[width=0.58\linewidth]{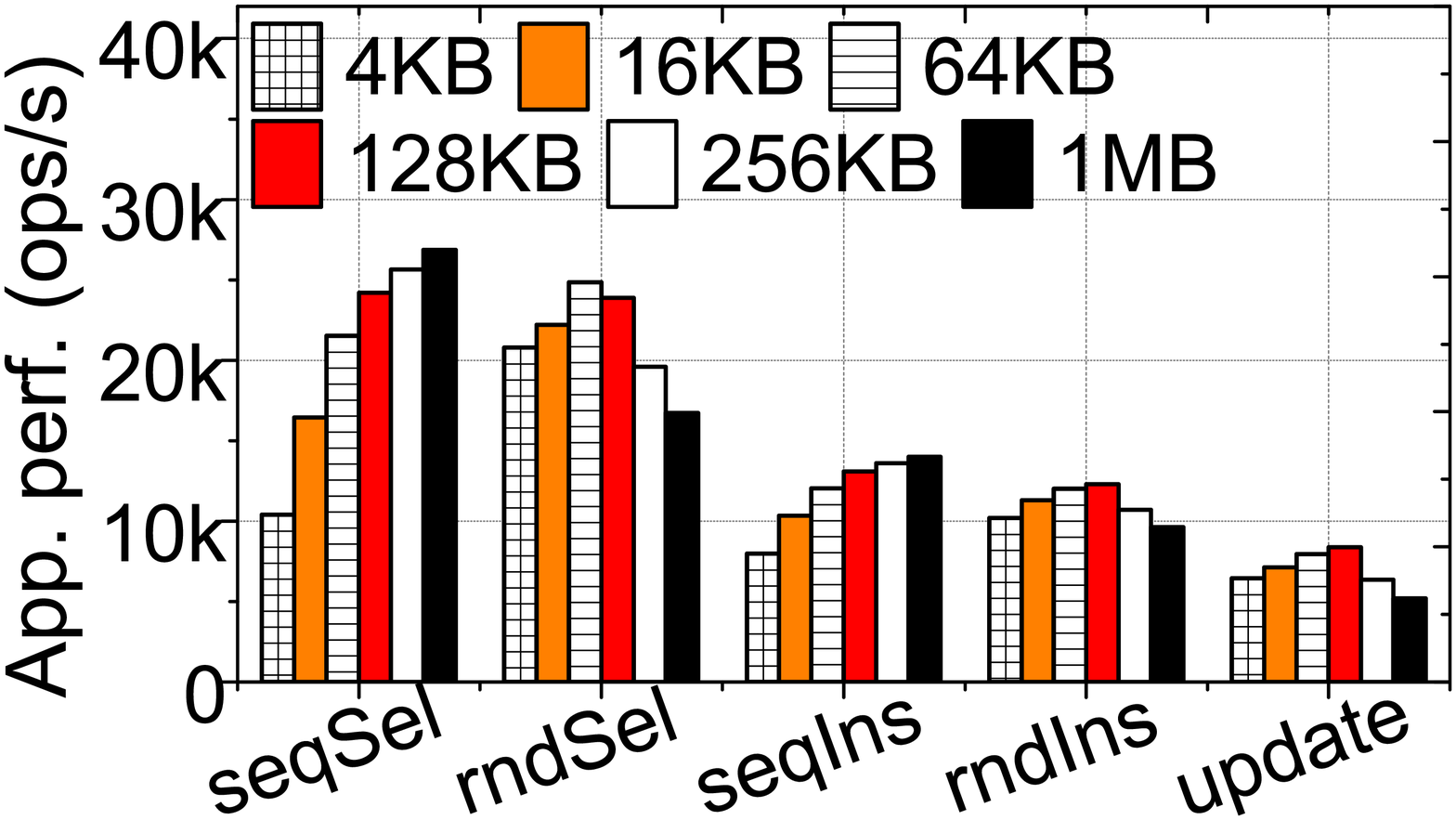}}}
\subfloat[Larger footprints.]{\label{fig:datasize}\rotatebox{0}{\includegraphics[width=0.42\linewidth]{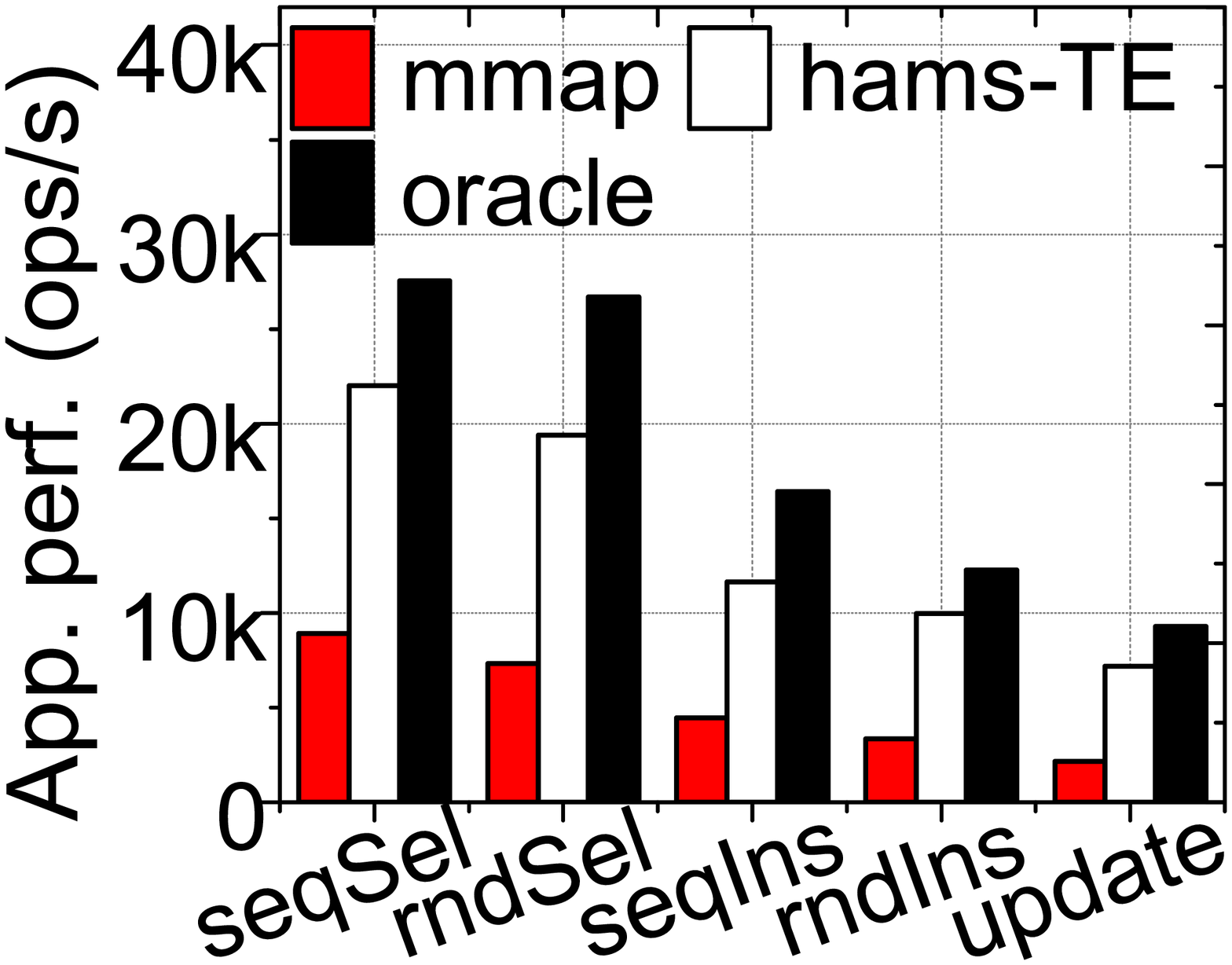}}}
\caption{\label{fig:sensitive_test}Performance impact of different page sizes and large memory footprints.\vspace{-5pt}}
\vspace{-15pt}
\end{figure}

\newedit{
\subsection{Sensitive Testing}
\noindent \textbf{Various page sizes.} We evaluate the performance of SQLite benchmark with various page sizes, and the results are shown in Figure \ref{fig:pagesize}. While 4KB and 1MB are the default page sizes in Linux Kernel 4.9, we also select intermediate page sizes such as 16KB, 64KB, 128KB and 256KB. As a small page size incurs frequent TLB misses and cannot utilize ULL-Flash internal parallelism, \texttt{4KB} achieves poor performance in workloads \emph{seqSel} and \emph{seqIns}. On the other hand, employing a large page size increases data migration overhead when cache misses in NVDIMM. Therefore, \texttt{1MB} achieves poor performance in workloads of random accesses (e.g., \emph{rndSel} and \emph{rndIns}). In our evaluation, configuring the page size as \texttt{128KB} can achieve the best performance in most workloads.

\noindent \textbf{Large memory footprints.} We perform a stress testing on NVDIMM by increasing the data set size to 44 GB, and the results are shown in Figure \ref{fig:datasize}. \texttt{hams-TE} degrades the performance by 24\% compared to \texttt{Oracle}, owing to the frequent data migration between NVDIMM and ULL-Flash. Nevertheless, \texttt{hams-TE} still outperforms \texttt{mmap} by 181\%.
}

\section{Related Work and Discussion}
\label{sec:relatedwork}
Recently, Intel has released a byte-addressable PRAM-based NVDIMM (i.e., Optane DC PMM) as a replacement for the main memory \cite{optaneDC}. 
However, unlike DRAM, user applications \emph{cannot} directly access persistent data from the proposed hardware using load/store instructions without customized software stack. Specifically, OS needs a series of Intel-custom software support, including a block driver, persistent-memory-aware filesystem and \emph{Direct Access} (i.e., DAX) \cite{optane-sw} to directly map the Optane DC PMM (as memory) to a userspace. The existing applications also require modifications to be compatible with Intel runtime libraries \cite{pmdk} built on DAX.
Further, Optane DC PMM also has several drawbacks from a hardware design angle. Specifically, it exhibits much lower storage capacity than ULL-Flash based HAMS (i.e., 512 GB/DIMM vs. 2.3 TB/DIMM) \cite{optane-size}. With the same number of memory packages per standard unit size, the aggregated throughput of Optane DC PMM is 4.5$\times$ lower than that of ULL-Flash \cite{zng2020,optane-perf}. 
It also faces the challenges of addressing the long PRAM write latency issues. 
\cite{yang2020empirical} reports that Optane DC PMM integrates a 16KB XPBuffer to accommodate the write requests. However, as the XPBuffer size is fixed and relatively small, \cite{optane-perf} observes that NVDIMM-N outperforms Optane DC PMM (as persistent memory) by 5.72x in write-intensive workloads. 



\newedit{Several prior studies \cite{hybridimm,lee2020nvdimm, jayaraj2015potential, voskuilen2016analyzing, hammond2016multi} propose to integrate DRAM and flash into a single system memory.} Similar to Optane DC PMM, memory requests need to go through multiple software layers, including NVML libraries and a specific HybriDIMM driver \cite{hybridimm-demo}, before accessing data from HybriDIMM. In addition, when configuring HybriDIMM as persistent memory, its internal DRAM buffer is \textit{disabled}, which directly exposes the long flash latency to system \cite{hybridimm-brief}.

Abulila \textit{et al.} proposes FlatFlash \cite{abulila2019flatflash}, which utilizes NAND flash to expand the memory space. Specifically, FlatFlash directly exposes ULL-Flash to the host as a byte-addressable device by leveraging the SSD internal DRAM as cache. However, a large portion of the SSD internal DRAM is used to store the address translation table \cite{zhang2019flashgpu}. The remaining DRAM space is much smaller than the host-side DRAM, which can be insufficient to accommodate the whole working set. In addition, as FlatFlash employs MMIO rather NVMe protocol to access the underlying ULL-Flash, it cannot benefit from the SSD internal parallelism thereby exhibiting lower device-level throughput. While migrating hot pages to the host-side memory can improve the overall performance, FlatFlash cannot guarantee the data persistency in such case.

In contrast, our HAMS expands the capacity of main memory without modifying the traditional filesystem or user applications. Specifically, just like DRAM, it directly exposes the address space of ULL-Flash to MMU, while leveraging our HAMS controller to manage data movements between NVDIMM and ULL-Flash, making it transparent to OS. 
To the best of our knowledge, such architectural design has not been discussed in the literature before. While HAMS can also be implemented as a kernel module, it requires OS to respond to every cache miss in NVDIMM (i.e., page fault), which incurs the overhead of context switch and page fault handling. Such software overhead is undesirable when large working sets incur frequent page swapping between NVDIMM and ULL-Flash (cf. Figure \ref{fig:oval_motiv1}). 
Furthermore, HAMS outperforms other DRAM+NVM approaches by maximizing the throughput of both NVDIMM and ULL-Flash (cf. Section \ref{sec:performance_analysis}). Our persistency control design can also guarantee data persistency without sacrificing ULL-Flash's performance. 

\newedit{
A set of prior work propose disaggregated memory solutions to expand the memory capacity \cite{ousterhout2010case, lim2009disaggregated, klimovic2017reflex, kommareddy2019page}. For example, \cite{ousterhout2010case} explores the feasibility of constructing a large memory pool across 1,000 servers via Ethernet. However, this design suffers from a low network bandwidth and high total cost of ownership (TCO). 
\cite{lim2009disaggregated} partially addresses the aforementioned challenges. \cite{lim2009disaggregated} improves the network throughput by employing PCIe interface and reduces the cost of hardware infrastructures by deploying more DRAM DIMMs in customized blade servers. Unfortunately, it is still challenging to adopt this design, owing to the high cost (i.e., price and power consumption) of DRAM DIMMs. 
\cite{klimovic2017reflex} further reduces TCO by replacing DRAM with NAND flash. However, accessing flash from remote servers increases the I/O latency by 10$\sim$15 us, which is 5$\times$ longer than ZNAND access latency (i.e., 3us). On the other hand, \cite{klimovic2017reflex} requires source-level modifications to the running applications, which exposes huge overheads to the users. 
In contrast to the above solutions, HAMS is a scale-up solution, which aggregates the capacities of local NVDIMM and ULL-Flash as a single memory space. HAMS, therefore, saves the huge cost of constructing many blade servers and purchasing expensive DRAM DIMMs. In addition, as HAMS builds TB-scale persistent memory in an OS-transparent manner, executing applications in HAMS requires no changes to the existing programming models. 
}


\section{Conclusion}
\label{sec:conclusion}
We proposed HAMS to aggregate the storage capacities of NVDIMM and ULL-Flash into a single large memory space, which can be used either as a working memory expansion or as a persistent memory expansion. We also optimized HAMS by modifying its datapath and hardware modules, which guarantees data persistency and makes HAMS more energy efficient and reliable. 
Our HAMS and advanced HAMS architectures improve MIPS by 97\% and 119\%, respectively, compared to the software-based hybrid NVDIMM design, while saving 41\% and 45\% energy, respectively.

\section{Acknowledgement}
This research is mainly supported by NRF 2021R1AC4001773 and IITP 2021-0-00524. The work is also supported in part by KAIST start-up package (G01190015), NRF 2016R1C182015312, and MemRay grant (G01190170). Dr. Kandemir is supported in part by NSF grants 1908793, 1629129, 2028929, and 1931531. Other product names used in this publication are for identification purposes only and may be trademarks of their respective companies. Myoungsoo Jung is the corresponding author.


\bibliographystyle{IEEEtranS}
\bibliography{ref}

\end{document}